\documentclass[aps,prx,twocolumn,reprint,floatfix, superscriptaddress]{revtex4-2}

\usepackage{leftindex}
\usepackage[normalem]{ulem}
\usepackage[utf8]{inputenc}
\usepackage{graphicx,amsmath,amssymb,braket}
\usepackage{color}

\definecolor{darkgreen}{rgb}{0.0, 0.4, 0.26}

\usepackage[colorlinks=true,citecolor=myblue,linkcolor=myred]{hyperref}

\definecolor{mygrey}{gray}{0.35}
\definecolor{myblue}{rgb}{0.2,0.2,0.8}
\definecolor{myzard}{cmyk}{0,0,0.05,0}
\definecolor{mywhite}{rgb}{1,1,1}
\definecolor{mywhite}{rgb}{1,1,1}
\definecolor{myred}{rgb}{1,0.,0.3}

\def\be{\begin{equation}}
\def\ee{\end{equation}}
\def\ba{\begin{align}}
\def\enda{\end{align}}
\def\bi{\begin{itemize}}
\def\ei{\end{itemize}}

\def\beq{\begin{equation}}
\def\beq{\begin{equation}}
\def\eeq{\end{equation}}

\begin{document}

\title{Passive photonic CZ gate with two-level emitters in chiral multi-mode waveguide QED}

\author{Tom\'as Levy-Yeyati}
\email{tomas.levy@iff.csic.es}
\affiliation{Institute of Fundamental Physics IFF-CSIC, Calle Serrano 113b, 28006 Madrid, Spain.}
\author{Carlos Vega}
\affiliation{Institute of Fundamental Physics IFF-CSIC, Calle Serrano 113b, 28006 Madrid, Spain.}
\author{Tom\'as Ramos}
\affiliation{Institute of Fundamental Physics IFF-CSIC, Calle Serrano 113b, 28006 Madrid, Spain.}
\author{Alejandro Gonz\'alez-Tudela}
\email{a.gonzalez-tudela@csic.es}
\affiliation{Institute of Fundamental Physics IFF-CSIC, Calle Serrano 113b, 28006 Madrid, Spain.}

\begin{abstract}
Engineering deterministic photonic gates with simple resources is one of the long-standing challenges in photonic quantum computing. Here, we design a passive conditional gate between co-propagating photons using an array of only two-level emitters. The key resource is to harness the effective photon-photon interaction induced by the chiral coupling of the emitter array to two waveguide modes with different resonant momenta at the emitter's transition frequency. By studying the system's multi-photon scattering response, we demonstrate that, in certain limits, this configuration induces a non-linear $\pi$-phase shift between the polariton eigenstates of the system without distorting spectrally the wavepackets. Then, we show how to harness this non-linear phase shift to engineer a conditional, deterministic photonic gate in different qubit encodings, with a fidelity  arbitrarily close to 1 in the limit of large number of emitters and coupling efficiency. Our configuration can be implemented in topological photonic platforms with multiple chiral edge modes, opening their use for quantum information processing, or in other setups where such chiral multi-mode waveguide scenario can be obtained, e.g., in spin-orbit coupled optical fibers or photonic crystal waveguides.
\end{abstract}

\maketitle

\section{Introduction}~\label{sec:intro}

\begin{figure}[!ht]
    \centering
    \includegraphics[width=0.85\columnwidth]{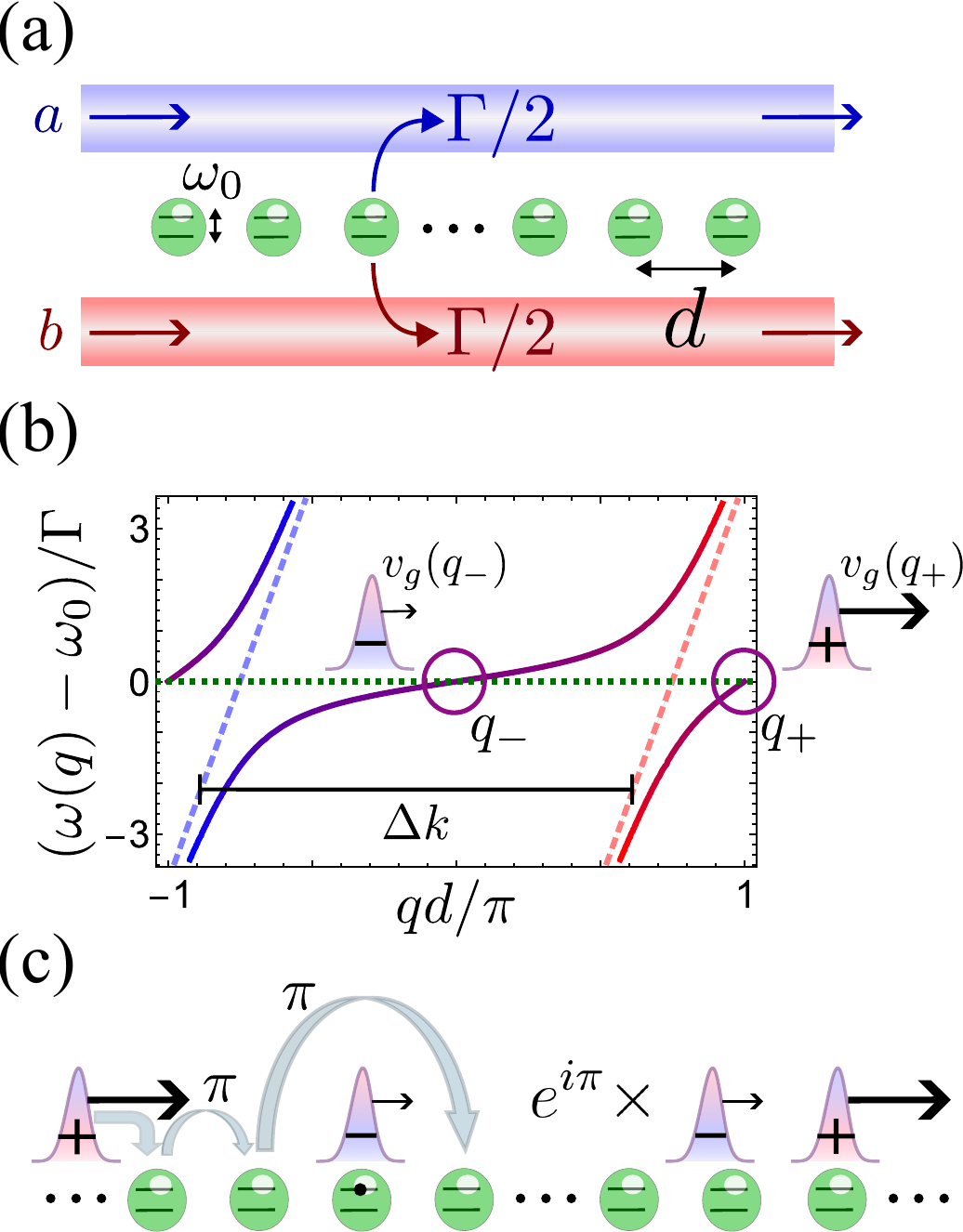}
    \caption{(a) Schematic representation of the chiral multimode waveguide QED setup: a periodic array of two-level emitters chirally coupled to two waveguides, $a$ and $b$. (b) In dashed lines, energy dispersion of the uncoupled system: in blue (red) the $a$ ($b$) photonic modes, in green the emitters' resonance. For the same frequency, photons propagating in different
    channels have a difference in momentum $\Delta k$. In solid lines, polariton dispersion of the coupled system. The two resonant excitations of the array are symmetric/antisymmetric  ($+$/$-$) superpositions of the original photonic modes: they have different group velocities and acquire, respectively, a $\pi$/$0$-phase for each emitter they encounter. (c) Schematic representation of the scattering of the two resonant polaritons: the fastest one can overcome the slowest and acquire a non-linear $\pi$-phase shift due to the emitters' saturation.}
    \label{fig:1}
\end{figure}

Engineering strong quantum correlations at the single photon level is an outstanding challenge due to the weakly interacting character of the photons~\cite{Chang2014}. One of the ultimate goals of this field is the generation of deterministic conditional gates between photons since, combined with arbitrary single photon operations, they can generate arbitrary quantum states of light. Some proposed methods use photon-storage protocols in different media~\cite{masalas2004,friedler2005,andre2005,gorshkov10a,gorshkov2011,thompson2017,Ioakoupov2018,das2016} to induce a conditional phase in a traveling photon depending on the previous storage (or not) of another photon. Other methods avoid the need for storing and retrieving photons by making them overlap sending counter-propagating pulses~\cite{He2014,brod2016,brod2016a,Schrinski2022PassiveEmitters} through some non-linear medium. Ideally, one would like the non-linearity to be as simple as possible, such as a two-level emitter, that is typically available in all platforms ranging from the microwave to the optical regime. However, it is known that achieving large phase shifts with two-level systems usually comes at the price of spectral entanglement, spoiling the gate fidelity~\cite{Shapiro2006,GeaBanaloche2010,Nysteen2017}. For this reason, current proposals use more complicated non-linear media to circumvent this limitation, such as Rydberg atoms~\cite{masalas2004,friedler2005,andre2005,gorshkov10a,gorshkov2011,thompson2017,Ioakoupov2018,das2016,He2014}, cross-Kerr couplings~\cite{brod2016,brod2016a}, dynamical $\chi^{(2)}$ non-linearities~\cite{Heuck2020,Li2020a}, V-type emitters~\cite{Schrinski2022PassiveEmitters}, or quadratic light-matter interactions~\cite{Alushi2023WaveguideInteractions}, which might not be available in all platforms. 

\begin{figure*}
    \centering
   \includegraphics[width=0.85\textwidth]{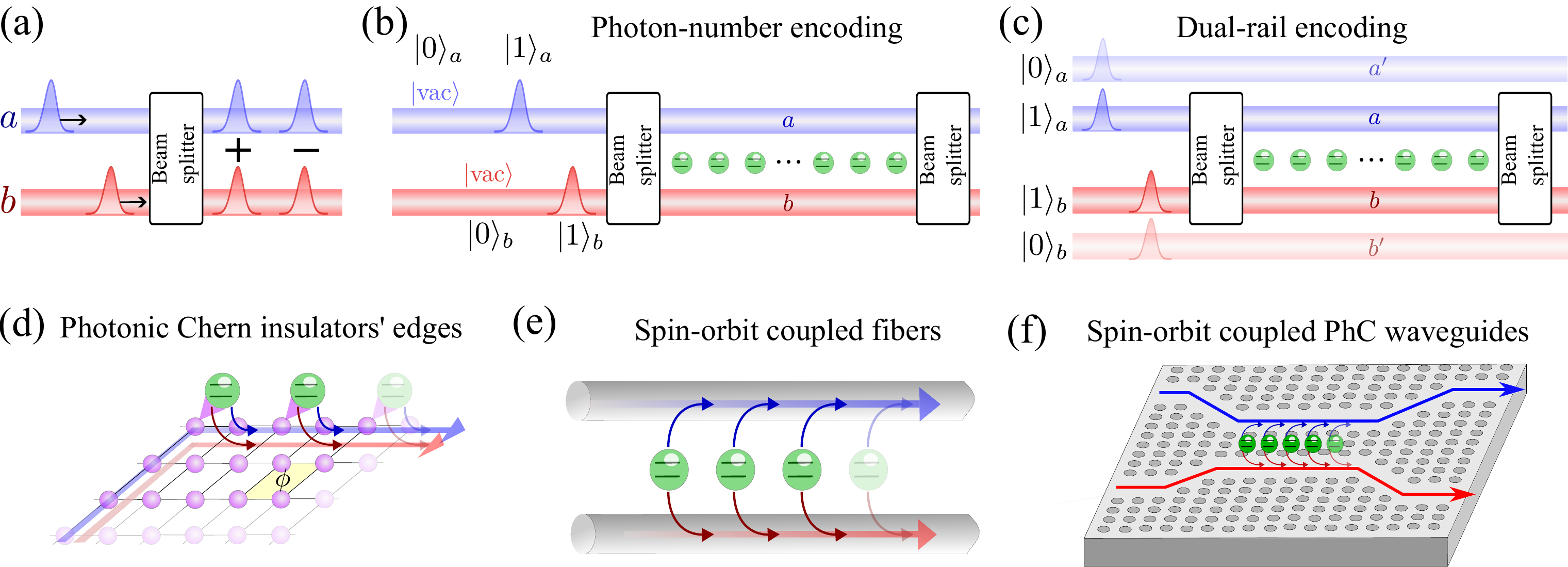}
    \caption{(a) Schematic representation of the beam splitter operation that transforms photons in the original channels into the eigenstates of the coupled chain of emitters, depicted in Fig.~\ref{fig:1}(a). (b-c) Possible photonic qubit encodings to harness the conditional $\pi$-shift depicted in Fig.~\ref{fig:1}(c): (b) a photon-number encoding, where the single qubit state is determined by the presence or absence of a photon; and (c) a dual-rail encoding, where the state is determined by the presence of a photon in the main (interacting) channel or in the auxiliary (non-interacting) channel. (d-f) Potential platforms to host the chiral, multi-mode waveguide QED setup: (d) emitters coupled to the edges of a topological photonic setup, with a band-structure with Chern number two; and chirally coupled emitters to (e) optical fibers or (f) photonic crystal waveguides. Note that these pictures are just a schematic representation of the different platforms, not actual designs. }
    \label{fig:Encodings}
\end{figure*}

In this work, we propose an alternative method to obtain a conditional gate between co-propagating photons based only on a periodic array of two-level emitters. The ingredients, intuition, and key results developed in this manuscript to design this photonic gate are summarized in Figs.~\ref{fig:1}-\ref{fig:Encodings}, and we elaborate them briefly in what follows.

\newpage

In the first part of the manuscript, we demonstrate the emergence of a different type of non-linear polariton interactions in the chiral multi-mode waveguide QED setup schematically depicted in Fig.~\ref{fig:1}(a). This setup consists of
a set of emitters, with transition frequency $\omega_0$ and equally spaced by a distance $d$, chirally coupled to two waveguide modes, $a$ and $b$.
A key difference of this setup, compared to previously considered coupled waveguide configurations~\cite{Schrinski2023}, is that the energy dispersion of the modes must be different, such that a photon on resonance with the emitters has a different momentum depending on the waveguide in which it is propagating. This introduces a new parameter, i.e., the multi-mode phase difference ($\Delta k d$), that affects the dispersion of the photons inside the emitters' medium, i.e., the polariton dispersion, resulting from the avoided crossing of the uncoupled waveguide and emitters' energy dispersion, shown in red/blue dashed and green dotted lines, respectively, in Fig.~\ref{fig:1}(b). In particular, in this manuscript we demonstrate that for any non-trivial $\Delta kd$, there are two resonant polaritonic excitations that are symmetric/antisymmetric superpositions of the original waveguide modes and, importantly, each has a different group velocity. Besides, we also show that the two polaritons acquire different phases as they scatter off an emitter, in particular, the (anti)symmetric acquires a $\pi$($0$)-phase. Combining both features, we also prove that the scattering between these two polaritons can result in a non-linear $\pi$-phase shift when the fastest polariton overcomes the slowest one while crossing the chain. The intuition is that the polaritons acquire a $\pi$-shift less compared to the linear case due to the saturation of the two-level emitters, as schematically depicted in Fig.~\ref{fig:1}(c). Besides, in the limit of large number of emitters, inelastic scattering events, where the frequency of each photon changes, are suppressed due to the conservation of quasimomentum associated to the periodicity of the array, resulting in a perfectly elastic non-linear phase shift.

In the second part of the manuscript, we show how to harness this non-linear phase shift to obtain a passive CZ gate in different photonic qubit encodings. For this purpose, we need a final ingredient, that is a way of mapping resonant input photons coming from the original $a$ ($b$) channels, in which we will encode our qubits, into the (anti)symmetric photonic superpositions, which are the ones that acquire the desired phases; and then transforming them back into the original channels after the interaction with the array of emitters. This can be done with the aid of a beam splitter, which operates as depicted in Fig.~\ref{fig:Encodings}(a). For the photons to cross inside the non-linear medium, the one that will map into the fastest polariton has to be delayed before entering the array. The qubit encoding can then be defined in at least two different ways: a photon-number encoding, in which the state of each subsystem $\alpha=\{a,b\}$, i.e., $\ket{1}_\alpha$ or $\ket{0}_\alpha$, is given by the presence or absence of a photon in that channel,  respectively, as depicted in Fig.~\ref{fig:Encodings}(b); or a dual-rail encoding in which the state is given by the presence of a photon in the main, interacting channel or in an auxiliary, non-interacting channel,
respectively, as depicted in Fig.~\ref{fig:Encodings}(c).  In both cases, the non-linear $\pi$-phase is conditional on the input $\ket{1}_a\otimes\ket{1}_b$, where two photons get to interact with the emitters' array; thus defining a CZ gate. 

The choice of the encoding will depend on the particular platform considered. In Figs.~\ref{fig:Encodings}(d-f) we illustrate three representative platforms where this setup can be engineered. For example, in Fig.~\ref{fig:Encodings}(d)  we represent a topological photonic implementation in which the photonic channels are the chiral edge modes appearing in two-dimensional topological photonic setups with large Chern number. These can be obtained in photonic crystals~\cite{Skirlo2014,Skirlo2015}, or in circuit QED~\cite{Owens2022} and matter-wave square lattices~\cite{aidelsburger13a,miyake13a,aidelsburger15a}, adjusting the value of the effective value flux $\phi$ per unit cell~\cite{Vega2023TopologicalQED}. In that platform, the photon number encoding might be the most suitable one, since the number of available photonic channels can be restrictive. In the other two setups,   Figs.~\ref{fig:Encodings}(e-f),  the photonic channels are implemented as optical fibers~\cite{mitsch14a,sayrin15a} and photonic-crystal waveguides~\cite{S_llner_2015,Coles2016,Mahmoodian2017,Barik2018,Siampour2023,Martin2023}. In these systems, photons propagate in both directions, however, exploiting spin-orbit light coupling~\cite{lodahl17a} one can make the emitters to couple only to the right (or left) propagating photons~\cite{mitsch14a,sayrin15a,S_llner_2015,Coles2016,Mahmoodian2017,Barik2018,Siampour2023,Martin2023}. In those cases, the dual rail encoding can be used if other auxiliary waveguide modes are available.

The manuscript is structured as follows: in Section~\ref{sec:setup}, we describe the general setup; in Sections~\ref{sec:single}-\ref{sec:two} we analyze the single and two photon scattering properties of the setup, which we use in Section~\ref{sec:gate} to design the gate architecture and calculate its fidelity. Finally, we summarize our findings in Section~\ref{sec:conclusions}.

\section{General setup~\label{sec:setup}}

In this manuscript, we take an agnostic perspective regarding the platform implementation, and make a general description of the setup as depicted in Fig.~\ref{fig:1}(a). The critical part of the setup consists in an array of $N$ two-level emitters, with transition frequency $\omega_0$ and separation $d$, that are chirally coupled with rates $\Gamma_\alpha$ to two waveguide modes with different energy dispersions, $\Omega_\alpha(k)$, as schematically depicted in Fig.~\ref{fig:1}(b) with dashed lines [we use $\alpha$ as the index referring to the waveguide channels $\alpha=\{a,b\}$]. In particular, it is important that the resonant momenta $(k_\alpha)$, defined by $\Omega_\alpha(k_\alpha)=\omega_0$, of each channel are different, such that $\Delta k d=(k_b-k_a)d$ generates an effective multi-mode phase difference when the photons propagate between the emitters within the different channels.

In addition, we consider that the $N$ emitters have a single transition between a ground ($g$) and excited ($e$) state described by a dipole operator $\hat{\sigma}=\ket{g}\bra{e}$ with frequency $\omega_0$ that we assume to be equal for all the emitters. Thus, they are described by the following Hamiltonian  (we use $\hbar\equiv 1$ along the manuscript):
\begin{align}
\hat{H}_S=\omega_0\sum_{n=1}^N\hat{\sigma}_n^\dagger \hat{\sigma}_{n}\,,\label{eq:HS}
\end{align}
where we use latin indices $n$ to denote the different emitters. The Hamiltonian describing the two-mode waveguide can be written in general as follows:
\begin{align}
\hat{H}_B=\sum_{\alpha}\int dk\ \hat{\alpha}_k^\dagger   \Omega_\alpha(k) \hat{\alpha}_k\,,\label{eq:HB_1}
\end{align}
where $\hat{\alpha}^{(\dagger)}_k$ is the destruction (creation) operator corresponding to the $\alpha$ photonic channel with associated momentum $k$ and energy $\Omega_\alpha(k)$. As depicted in Fig.~\ref{fig:1}(b), we take a simplified description of the waveguide modes by linearizing their energy dispersion around the atomic frequency:
\begin{align}
\Omega_\alpha(k)\approx \omega_0 + c_\alpha(k-k_\alpha)\,,\label{eq:disp}
\end{align}
where $c_\alpha$ is the effective light speed of the guided modes resonant to the emitter frequencies, and $k_\alpha$ their corresponding associated momenta. 

 We assume the coupling between the emitters and the photonic modes to be spatially local and excitation-conserving, which is generally a good approximation in most implementations. Therefore, the interaction part of the Hamiltonian can be generally written as:
\begin{align}
\hat{H}_\mathrm{int}=\sum_{n=1}^N\sum_{\alpha}\left(g_\alpha \hat{\alpha}^\dagger_{x_n}\hat{\sigma}_n+\mathrm{H.c.}\right)\,,\label{eq:HSint}
\end{align}
where we assume that all the emitters couple equally to the $\alpha$ mode with strength $g_\alpha$'s independent of their position, $x_n=nd$, and where $\hat{\alpha}^{\dagger}_{x}$ is the real-space Fourier transform of $\hat{\alpha}^\dagger_k$ defined as:
\begin{align}
\label{eq:fourier}
\hat{\alpha}^\dagger_x=\frac{1}{\sqrt{2\pi}}\int dk\ \hat{\alpha}^\dagger_k e^{-i k x}\,.
\end{align}

Thus, the complete light-matter Hamiltonian of the chiral multi-mode waveguide QED setup can be written as: $\hat{H}=\hat{H}_S+\hat{H}_B+\hat{H}_\mathrm{int}$. Making a unitary transformation that rotates with the frequency of the emitters, $\omega_0$, we can re-write the complete light-matter Hamiltonian in real space as:
\begin{align}
\hat{H}&=\sum_{\alpha}c_\alpha\int dx\ \hat{\alpha}_x^\dagger(-i\partial_x-k_\alpha) \hat{\alpha}_x\nonumber\\
&+\sum_{n=1}^N\sum_{\alpha}\sqrt{\Gamma_\alpha c_\alpha}\left(\hat{\alpha}^\dagger_{x_n}\hat{\sigma}_n+\mathrm{H.c.}\right)\,,\label{eq: complete4}
\end{align}
where we already used $\Gamma_\alpha=\frac{|g_\alpha|^2}{c_\alpha}$, that is the decay rate of the emitters into each corresponding photonic channel. Although it is not strictly needed for engineering the conditional phase between the photons, it is practical to choose a situation where $c_a=c_b=c$ such that, after scattering with the emitters, the photons at the two channels propagate at the same speed. 

Assuming that the coupling is weak enough and neglecting the finite propagating time of the photons between the emitters (Born-Markov conditions), one can adiabatically eliminate the photonic field resulting in an effective master equation for the density matrix associated to the emitters' degrees of freedom that reads:
\begin{align}
\frac{\partial \rho}{\partial t}=-i\left(\hat{H}_\mathrm{eff}\rho-\rho\hat{H}^\dagger_\mathrm{eff}\right)+J[\rho]\,,\label{eq:meq}
\end{align}
where $J[\rho]$ embeds the quantum jump terms, and $\hat{H}_\mathrm{eff}$ is an effective spin Hamiltonian, which for this multi-mode cascaded situation reads~\cite{Vega2024}:
\begin{align}
\nonumber
\hat{H}_\mathrm{eff}=&-i\sum_\alpha\sum_{n}\frac{\Gamma_\alpha}{2}\hat{\sigma}_{n}^\dagger\hat{\sigma}_n\\
&-i  \sum_{\alpha}\sum_{n>m} \Gamma_\alpha e^{ik_  
 \alpha(x_n-x_m)}\hat\sigma^\dagger_n \hat \sigma_m\,.\label{eq:spinmodel}
\end{align}
The summation of the second term on the right-hand side of the equation takes into the account the chiral character of the modes. Differently from the standard chiral, single-mode~\cite{lodahl17a} or other coupled waveguide configurations~\cite{Schrinski2023} where the phases $e^{i k_\alpha x_n}$ can be gauged out, here, the relative position between the emitters matter because $k_\alpha d\neq k_\beta d$, which can give rise to qualitatively different phenomena. As we will see afterwards, this difference is critical for obtaining the conditional phase with only two-level emitter non-linearities. Finally, we will also phenomenologically include losses into other decay channels at rate $\gamma$ by adding an imaginary energy term into the emitters $\omega_0\rightarrow \omega_0-i\frac{\gamma}{2}$.

\section{Single-photon scattering properties~\label{sec:single}}

In this section, we derive the single-photon scattering properties of the system. For that, we follow two alternative approaches: first, in Section~\ref{subsec:effectivespineigenstates} we use the spin model derived in Eq.~\eqref{eq:spinmodel} to characterize the dispersion of the polaritonic (i.e., light-matter) excitations of the array in the limit of $N\rightarrow \infty$. Then, in Section~\ref{subsec:eigenstatesscattering}, we use the complete light-matter Hamiltonian of Eq.~\eqref{eq: complete4} to calculate the exact single-photon S-matrix. With that, we will characterize the photonic states associated with the different polariton modes, and consider the effects of finite pulse-widths and finite number of emitters $N$ in the chain.

\subsection{Using the effective spin model~\label{subsec:effectivespineigenstates}}

\begin{figure}[tb]
    \centering
    \includegraphics[width=0.8\columnwidth]{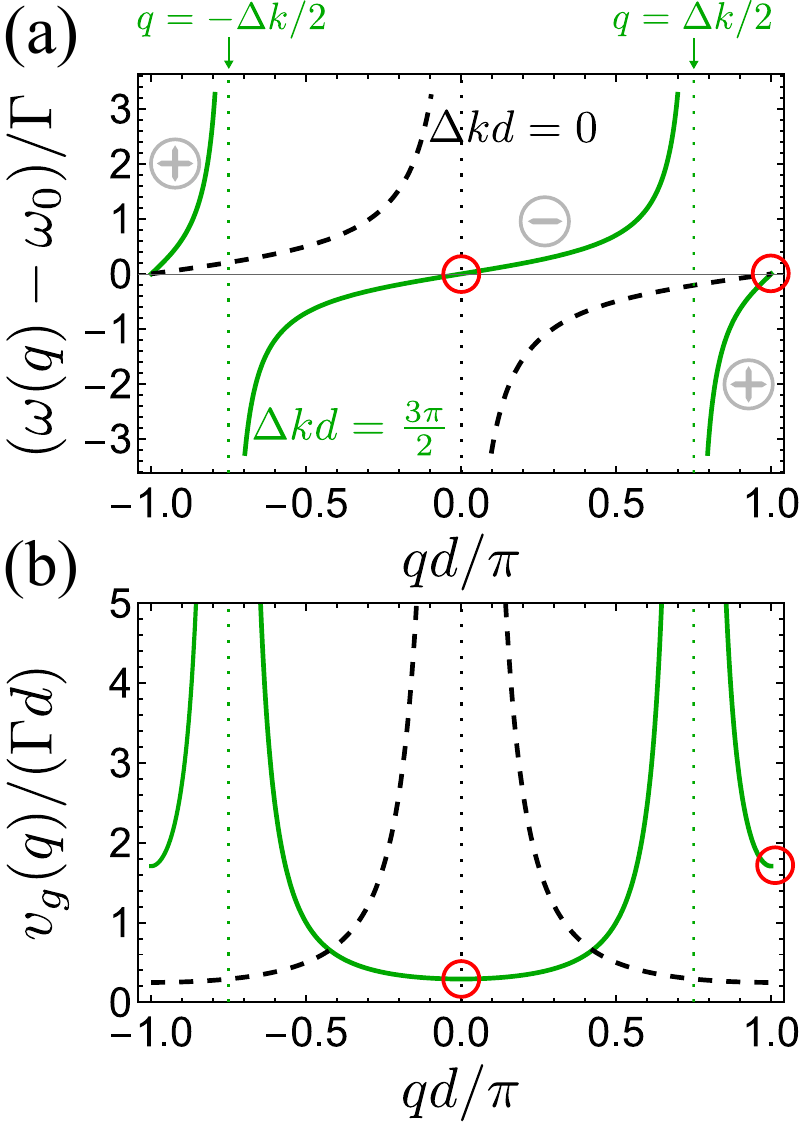}
    \caption{(a) Polariton energy dispersion $\omega(q)$ as defined in Eq.~\eqref{eq:wq} for $\Gamma_a=\Gamma_b=\Gamma/2$ and $\Delta k d=0$ ($ 3\pi/2$) in dashed black (solid green). When $\Delta k d\neq 0$, two bands appear that we label as $+$ and $-$ bands. Red circles indicate the two resonant momenta $qd=0, \pi$ when $\Delta k d\neq 0$. (b) Corresponding group velocity $v_g(q)$ for the same parameters as panel (a).  }
    \label{fig:2}
\end{figure}

For $N\rightarrow \infty$, the edges of the emitter ensemble do not play a role and we can diagonalize the single-excitation sector, i.e., $\sigma_n^\dagger\ket{g^{\otimes N}}$, using a plane-wave expansion of the operators. In particular, the effective spin Hamiltonian in the single excitation sector, denoted by $\hat H_{\mathrm{eff},\mathrm{1}}$, can be diagonalized as follows:
\begin{align}
    \hat H_{\mathrm{eff},\mathrm{1}}=\sum_q \omega(q)\ket{q}\bra{q}\,,
\end{align}
where $\ket{q}$ are elementary plane-wave excitations of the spin Hamiltonian defined by:
\begin{align}
    \ket{q}=\frac{1}{\sqrt{N}}\sum_n e^{i (k_0+q) x_n}\hat\sigma^\dagger_n\ket{g^{\otimes N}}\,,\label{eq:plane}
\end{align}
where $q\in[-\pi/d,\pi/d)$ is the quasimomentum associated to the polariton defined with respect to the mean of the two resonant momenta $k_\alpha$, i.e., $k_0=(k_a+k_b)/2$.

Despite the plane-wave eigenstates are defined using only spin operators, they represent hybrid light-matter excitations, i.e., polaritons, since their energies $\omega(q)$ are affected by the photon-mediated interactions of Eq.~\eqref{eq:spinmodel}. Concretely, their eigenenergies for the two-mode situation we are considering read:
\begin{align}
\label{eq:wq}
\omega(q)=\omega_0-\frac{\Gamma_a}{2}\cot{\left[\frac{q+\frac{\Delta k}{2}}{2}d\right]}-\frac{\Gamma_b}{2}\cot{\left[\frac{q-\frac{\Delta k}{2}}{2}d\right]}\,.
\end{align}

In Fig.~\ref{fig:2}(a) we plot this energy dispersion $\omega(q)$ for $\Gamma_a=\Gamma_b$ and two different $\Delta k d=(k_b-k_a)d$ values. In dashed black we plot the case $\Delta k d=0$ where, as expected, we recover the polariton dispersion of a single-mode chiral waveguide, like in Ref.~\cite{Schrinski2022PassiveEmitters}, with a decay rate $\Gamma=\Gamma_a+\Gamma_b$. However, when $\Delta k d\neq 0$ two interesting differences arise:
\begin{itemize}
    \item The spectrum has now double degeneracy, exhibiting two polariton bands: One in the middle of the first Brillouin zone (1BZ) and one on the edge of the 1BZ, that we label as $-$ and $+$ bands, respectively. This labelling is related to the photonic component of the polaritons, that we introduce in Section~\ref{sec:intro}, and will be discussed thoroughly in Section~\ref{subsec:eigenstatesscattering}. For $\Gamma_a=\Gamma_b$, on resonance, $\omega(q)=\omega_0$, the $-$ band hosts a polariton with $q_-=0$ and the $+$ band one with $q_+=\pi/d$, indicated with red circles in Fig.~\ref{fig:2}(a). Note the difference with the $\Delta k d=0$ case where there is only a single resonant momenta at $q=\pi/d$.

    \item 
    In Fig.~\ref{fig:2}(b), we plot the associated group velocity, $v_g(q)=\partial_q\omega(q)$, for the whole 1BZ, with $\Gamma_a=\Gamma_b=\Gamma/2$. Importantly, the group velocities associated with the two resonant polaritons,
\begin{align}
    \label{eq:vqresonant}
    v_g(0)=\frac{\Gamma d}{4\sin^2\left(\frac{\Delta kd}{4}\right)}, \hspace{3mm} v_g(\pi/d)=\frac{\Gamma d}{4\cos^2\left(\frac{\Delta kd}{4}\right)}\ ,
\end{align}
when $\Delta k d\neq n\pi$, with $n\in\mathbb{Z}$, are substantially different: $v_g(0) < v_g(\pi/d)$, for $\pi<\Delta k d<2\pi$ and $v_g(0) > v_g(\pi/d)$ for $0<\Delta k d<\pi$. This enables to obtain situations in which the faster polariton overcomes the slower one if the ensemble is large enough, as depicted in Fig.~\ref{fig:1}(c). 

\end{itemize}

In the next section, we use an exact scattering matrix calculation for finite $N$ to understand better the photonic nature of these different polariton eigenstates.

\begin{figure*}[tb]
    \centering
    \includegraphics[width=0.9\textwidth]{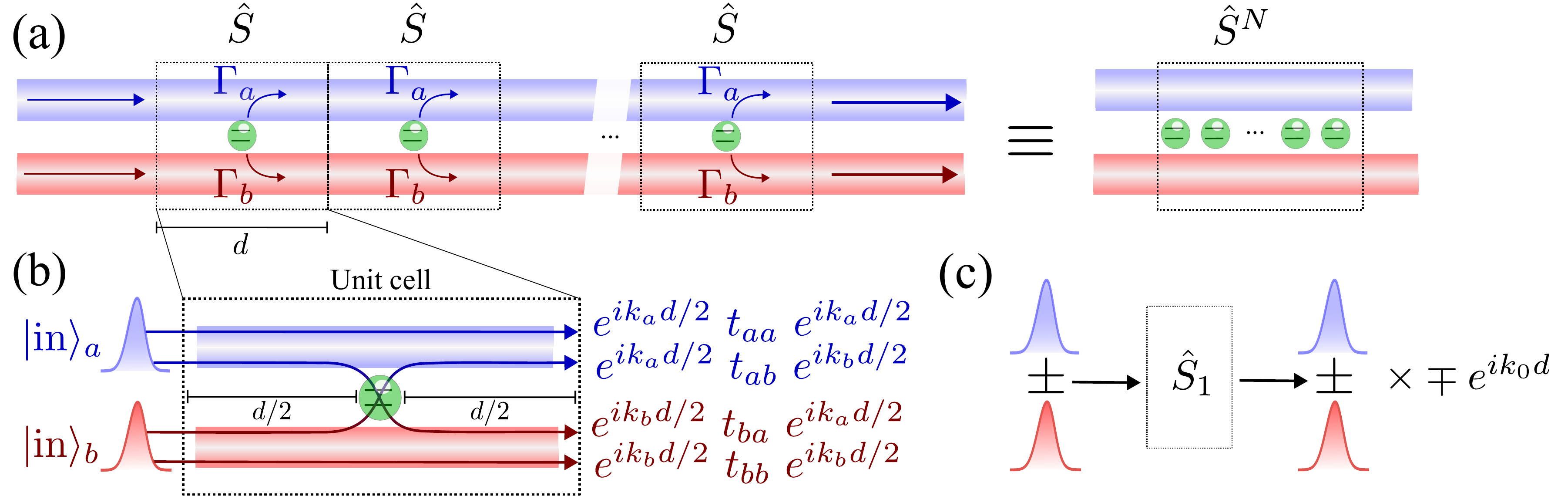}
    \caption{(a) Schematic representation of the photon scattering due to $N$ two-level emitters coupled to a two-mode chiral waveguide. Due to the chiral character of the two modes, the global S-matrix of the system can be calculated as a product of the $\hat{S}$ matrices of each unit cell. (b) Elements of the single photon scattering matrix of a unit cell, that we denote by $\hat{S}_1$: when a photon travels in the $a$ ($b$) mode, it can either remain propagating within the same channel after scattering with the emitter, with probability amplitude $t_{aa (bb)}$, or it can scatter into the other channel, with probability amplitude $t_{ba (ab)}$. Importantly, before  and after scattering with the emitter, the photons acquire a different propagation phase, $e^{ik_\alpha d/2}$, depending on the channels in which they propagate and which can not be neglected. (c) The eigenstates of the single-photon scattering matrix, $\hat{S}_1$, are linear superpositions between the $a$ and $b$ channels. When the photons are on resonance with $\omega_0$ and for $\Gamma_a=\Gamma_b$, they acquire both a global propagating phase $\mp e^{i k_0 d}$ after scattering, being $k_0=(k_a+k_b)/2$, where the sign depends on the symmetric/antisymmetric nature of the superposition. Importantly, these superpositions correspond to the $q d=0,\pi$ polariton modes with different group velocities that we find with the effective spin model in Sec.~\ref{subsec:effectivespineigenstates}.}
    \label{fig:3}
\end{figure*}

\subsection{Using the S-matrix~\label{subsec:eigenstatesscattering}}

The S-matrix defines the mapping between a free photon input state to a free photon output state after interacting with our chiral, multi-mode waveguide QED system, i.e., $\ket{\mathrm{out}}=\hat{S}_\text{tot}\ket{\mathrm{in}}$. Let us now make several remarks about notation and simplifications that can be made in our setup:
\begin{itemize}

\item We label as $\hat{S}$ the S-matrix operator corresponding to a unit cell of the array. Due to the periodic arrangement of the emitters and the chiral character of the photonic modes, we can calculate the global scattering matrix of the $N$ emitters, $\hat{S}_\text{tot}$, as a product of the S-matrices of the unit cell~\cite{Mahmoodian2020}:
\begin{equation}
    \hat{S}_\text{tot}=\hat{S}^N\ ,
\end{equation}
as represented in Fig.~\ref{fig:3}(a).

\item The unit cell of the array consists of a single emitter, placed in the middle, and the free (non-trivial) propagation over half unit cell distance $d/2$, before and after scattering with the emitter. Thus, the S-matrix can be written as:
\begin{equation}
    \label{eq:Sunitcell}
    \hat S = \hat P\ \hat s\ \hat P\ ,
\end{equation}
where $\hat s$ characterizes the emitter scattering component and $\hat P$ the photon propagation over half a unit cell distance $d/2$.

\item Since the total light-matter Hamiltonian conserves the number of excitations, we can calculate the $\hat{S}$ matrix operator in each $n$-excitation manifold~\cite{Caneva2015}, $\hat{S}_n$. In this section, we focus on the single-photon excitation manifold, i.e.,  $\hat{S}_1$.
\end{itemize}

As schematically depicted in Fig.~\ref{fig:3}(b), four different processes can occur when a photon of energy $\omega$ impinges into a unit cell, namely: 
\begin{itemize}
    \item A photon propagating in the $a$ ($b$) channel scatters with the emitter and continues propagating in the same channel with probability amplitude $t_{aa(bb)}(\omega)$.

    \item A photon propagating in the $a$ ($b$) channel changes into the orthogonal one, $b$ ($a$), with a probability amplitude $t_{ba(ab)}(\omega)$.

    \item Besides, an important aspect that must be taken into account in our setup is that, for the same photon energy, $\omega$, photons acquire different propagating phases, $e^{i k_\alpha(\omega) d/2}$, before and after scattering with the emitter, depending on the channels they propagate into,  $\alpha=\{a,b\}$. Here, $k_\alpha(\omega)=(\omega-\omega_0)/c_\alpha+k_\alpha$ is the inverse dispersion relation of $\Omega_\alpha(k)\equiv \omega $ defined in Eq.~\eqref{eq:disp}. 
\end{itemize}

Considering all these processes, the unit cell S-matrix components for a single photon read:

\begin{align}
    \nonumber
    \hat s_1= \sum_{\alpha,\beta} \int d\omega\  t_{\beta\alpha}(\omega)   \ket{\omega}_\beta\bra{\omega}_\alpha\hspace{5mm} ,\\
    \hspace{5mm}  \hat P_1= \sum_{\alpha} \int d\omega\ e^{
    ik_\alpha(\omega)d/2}\ket{\omega}_\alpha\bra{\omega}_\alpha\ , 
\end{align}
for $\alpha,\beta=\{a,b\}$, and $\ket{\omega}_\alpha$ being the state describing a plane-wave photon of frequency $\omega$ propagating in the $\alpha$ mode:
\begin{align}
    \ket{\omega}_\mathrm{\alpha}=\frac{1}{\sqrt{2\pi c_\alpha}}\int dx\ e^{i k_\alpha(\omega) x} \alpha_x^\dagger\ket{\mathrm{vac}}\,.
\end{align} 
Note that these states $\ket{\omega}_\mathrm{\alpha}$ form an orthonormal complete basis for the single-photon manifold and thus satisfy $ \leftindex_\beta {\braket{\omega'|\omega}}_\alpha=\delta_{\alpha\beta}\delta(\omega-\omega')$. 

Since the single-photon scattering process is elastic, we can express the single-photon S-matrix as a $2\times 2$ matrix for each frequency,
$\hat S_1(\omega)$, writing the $a$ and $b$ components of the field as a column vector:
\begin{equation}
    \label{eq:S1Matrix}
\hat S_1(\omega)=\hat P_1(\omega)\ \hat s_1(\omega)\ \hat  P_1(\omega)\ ,
\end{equation}
 where
\begin{align}
\label{eq:s1Matrix}
\nonumber
\hat{P}_1(\omega)&=\begin{pmatrix}
 e^{i k_a(\omega)d/2}  & 0 \\
0 & e^{i k_b(\omega) d/2} 
\end{pmatrix}\, , \\
\hat {s}_1(\omega)&=\begin{pmatrix}
t_{aa}(\omega) &  t_{ab}(\omega)  \\
 t_{ba}(\omega)  &  t_{bb}(\omega)  
\end{pmatrix}\ .
\end{align}

There are several approaches in the literature to calculate such scattering coefficients~\cite{Ramos2017,Caneva2015,shi15a,Zheng2010,Shen2007,Shen2007StronglyImpurity,Gonzalez-Ballestero2016}. In our case, we use the real space representation of the Hamiltonian $\hat{H}$, see Eq.~\eqref{eq: complete4}, for a single emitter and solve the time-independent Schrödinger equation~\cite{Zheng2010,Shen2007,Shen2007StronglyImpurity,Gonzalez-Ballestero2016}:
\begin{align}
    \label{eq:schreq}
    \hat{H}\ket{\Psi_1}=\omega\ket{\Psi_1}\,,
\end{align}
using the following wavefunction ansatz for the single-excitation subspace:
\begin{align}
    \nonumber
    &\ket{\Psi_1}=\\
    &\left(C_e \sigma^\dagger+\int dx\ \phi_a(x) \hat a_x^\dagger + \int dx\ \phi_b(x) \hat b_x^\dagger \right)\ket{\mathrm{vac}}\otimes\ket{g}\,,
\end{align}
and for different boundary conditions for $\phi_\alpha(x<0)$, denoting as $0$ the emitter position. Using a plane wave within the $\alpha$-waveguide, i.e., $\phi_\alpha(x<0)=\frac{1}{\sqrt{2\pi c_\alpha}}e^{ik_\alpha(\omega) x}$, we can calculate the $t_{\alpha \beta}$ through the photon components at the other side of the emitter, i.e., $\phi_{\beta}(x>0)=\frac{1}{\sqrt{2\pi c_\beta}}t_{\beta \alpha}(\omega)e^{i k_\beta(\omega) x}$. Using these definitions, the solution of Eq.~\eqref{eq:schreq} leads to:
\begin{align}
    \nonumber
    t_{aa}(\omega)&=\frac{\gamma+\Gamma_b-\Gamma_a-2i(\omega-\omega_0)}{\gamma+\Gamma_a+\Gamma_b- 2i(\omega-\omega_0)}\ ,\\
    \nonumber
    t_{bb}(\omega)&=\frac{\gamma+\Gamma_a-\Gamma_b-2i(\omega-\omega_0)}{\gamma+\Gamma_a+\Gamma_b- 2i(\omega-\omega_0)}\ ,\\
    t_{ba}(\omega)=t_{ab}(\omega)&
    =\frac{-2\sqrt{\Gamma_a\Gamma_b}}{\gamma+\Gamma_a+\Gamma_b - 2i(\omega-\omega_0)}\ ,
\end{align}
where we also include the effect of the other decay channels at rate $\gamma$ \footnote{When $\gamma\equiv 0$, these coefficients satisfy the continuity equations of conservation of probability which imply: $|t_{aa}|^2+|t_{ba}|^2 =1$, $|t_{bb}|^2+|t_{ab}|^2=1$}. 

\begin{figure}[tb]
    \centering
    \includegraphics[width=0.9\columnwidth]{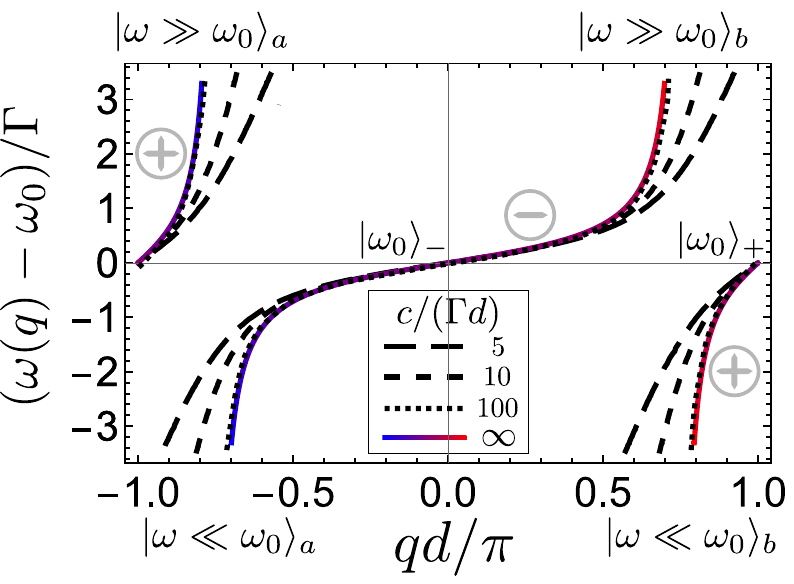}
    \caption{In dashed black lines, exact polariton energy dispersion $\omega(q)$ calculated by solving self-consistently Eq.~\eqref{eq:exactdisp}. Parameters are $\Delta k d=3\pi/2$, $\Gamma_a=\Gamma_b=\Gamma/2$, $c_a=c_b=c$, for different light speeds, as indicated in the legend. Solid line represents the calculation in the Markov limit, i.e., $c/(\Gamma d)\rightarrow \infty$. In color, representation of the type of scattering eigenstates associated to each polariton: blue represents photons in the $a$ mode; red, photons in the $b$ mode, and violet superpositions of $a$ and $b$ modes. Far from resonance, freely propagating $a$ or $b$ modes are eigenstates; near resonance, where the interaction with the emitters is noticeable, the eigenstates are superpositions of both modes. We highlight the resonant transfer eigenstates: $\ket{\omega_0}_-$, associated to the $q=0$ polariton, and $\ket{\omega_0}_+$, associated to the $q=\pi/d$ polariton.} 
    \label{fig:4}
\end{figure}

A relevant set of photonic states are those which remain unchanged, up to a phase, after scattering with the chain, i.e., the eigenvectors of the single-photon unit cell scattering matrix $\hat{S}_1(\omega)$, which we call the transfer eigenstates. These states satisfy:
\begin{align}
\hat{S}_1(\omega)\ket{\omega}_{\mu}=e^{i k_0 d}e^{iq_{\mu}(\omega)d}\ket{\omega}_{\mu}\,,\label{eq:scatsin}
\end{align}
where we use the index $\mu=\{+,-\}$ to denote each one of the transfer eigenstates for a given $\omega$ (two eigenvectors for a $2\times 2$ matrix), and $q_{\mu}(\omega)d$ denotes their corresponding phase acquired over a unit cell beyond the global propagating phase, $k_0 d$. The transfer eigenstates are, in general, weighted superpositions of $a$ and $b$ modes, and read:
\begin{align}\label{eqSM:eigenstates}
    \nonumber
    \ket{\omega}_+&=\sin \left(\frac{\theta_\omega}{2}\right) \ket{\omega}_a +  \cos \left(\frac{\theta_\omega}{2}\right) \ket{\omega}_b\,\\
    \ket{\omega}_- &=\cos \left(\frac{\theta_\omega}{2}\right) \ket{\omega}_a -  \sin \left(\frac{\theta_\omega}{2}\right) \ket{\omega}_b\ ,
\end{align}
where
\begin{align}
    \nonumber
    &\theta_\omega=\\\
    &\mathrm{arccot}\left [ \frac{\Gamma_b-\Gamma_a}{2\sqrt{\Gamma_a\Gamma_b}}\cos \left( \frac{\Delta kd}{2} \right)-\frac{\omega-\omega_0}{\sqrt{\Gamma_a\Gamma_b}}\sin \left( \frac{\Delta kd}{2} \right)  \right]\ .
\end{align}
The eigenvalue equation of Eq.~\eqref{eq:scatsin} leads to the following self-consistent equation:
\begin{align}
    \nonumber
    \omega-\omega_0=&-\frac{\Gamma_a}{2}\cot\left[\frac{q+\frac{\Delta k}{2}-\frac{\omega-\omega_0}{c_a}}{2}d\right]\\
    &-\frac{\Gamma_b}{2}\cot\left[\frac{q-\frac{\Delta k}{2}-\frac{\omega-\omega_0}{c_b}}{2}d\right]\,. \label{eq:exactdisp}
\end{align}
Solving such equation for every $q$, one arrives to the dispersion relation $\omega(q)$, that we plot in black dashed lines in Fig.~\ref{fig:4} for different $c_\alpha$. There are two bands, one in the middle of the 1BZ, the $-$ band, which gives the phase-shift $q_-(\omega)$ associated to the transfer eigenstate $\ket{\omega}_-$, and the $+$ band, in the edge of the 1BZ, giving the phase-shift $q_+(\omega)$ associated to $\ket{\omega}_+$.  

Remarkably, we can associate each transfer eigenstate with acquired phase $(k_0+q)d$ to a polariton of quasimomentum $k_0+q$, from Eq.~\eqref{eq:plane}. The exact dispersion of Eq.~\eqref{eq:exactdisp} is a non-Markovian version of the one obtained with the spin model, from Eq.~\eqref{eq:wq}, which considers the effect of retardation due to a finite light speed $c_\alpha$. In fact, in the limit $c_\alpha\gg\Gamma d$, we recover the Markovian dispersion, as shown in Fig.~\ref{fig:4} with the solid, colored line. The same Markovian limit can also be obtained by solving Eq.~\eqref{eq:scatsin} while approximating the free propagation phases $k_\alpha(\omega)$ from the S-matrix of Eq.~\eqref{eq:s1Matrix} to those on resonance $k_\alpha\equiv k_\alpha (\omega_0)$.

To understand better the nature of these eigenstates it is instructive to consider certain limits:

\begin{itemize}
\item When the photons are on resonance, i.e., $\omega=\omega_0$, and $\Gamma_a=\Gamma_b=\Gamma/2$, the transfer eigenstates are symmetric ($+$) and antisymmetric ($-$) superpositions of $a$ and $b$ modes, as in Fig.~\ref{fig:3}(c), and read:
\begin{align}
\label{eq:eigensreson}
\ket{\omega_0}_\pm=\frac{1}{\sqrt{2}}\left(\ket{\omega_0}_a\pm \ket{\omega_0}_b\right)\,,
\end{align}
with associated acquired phases over unit cell transfer: $q_{-}(\omega_0)d=0$ and $q_+(\omega_0)d=\pi$. From this, we can automatically get the phase acquired by the eigenstates when crossing the $N$ emitters, since $\hat S_1^N(\omega_0)\ket{\omega_0}_\pm=e^{iN(k_0+q_{\pm})d}\ket{\omega_0}_{\pm}$. Importantly, the $\ket{\omega_0}_{+}$ eigenstate behaves like the single-photon state in a single-mode chiral waveguide, that is, acquiring a $\pi$-shift after scattering with each emitter. The $\ket{\omega_0}_{-}$ eigenstate acquires no phase-shift after scattering with the emitters, however it does saturate their response. In addition to the described phase-shifts, a resonant finite-bandwidth transfer eigenstate pulse will suffer a different delay 
\begin{align}
    \label{eq:delays}
    \nonumber
    \tau_+&=\frac{d}{v_g(\pi/d)}=\frac{4}{\Gamma}\cos^2\left( \frac{\Delta kd }{4}\right)\ ,\\
    \tau_-&=\frac{d}{v_g(0)}=\frac{4}{\Gamma}\sin^2\left( \frac{\Delta kd }{4}\right)\ ,
\end{align}
for each emitter that it encounters, depending on the symmetric/antisymmetric character of the superposition, respectively.
This will be critical when engineering the gate, allowing for a non-linear $\pi$-phase shift as the photons cross inside the array in the two-photon regime, as shown in Fig.~\ref{fig:1}(c). 

\item On the contrary, when $|\omega-\omega_0|\gg \Gamma_{\alpha}$, then $q_{\mu}(\omega)=k_\alpha(\omega)-k_0$, and: 
\begin{align}
\ket{\omega\gg \omega_0}_{\mu}\approx \ket{\omega\gg\omega_0}_{a/b}\,,
\end{align}
that is, one recovers the free-propagating states. 
\end{itemize}
We encode the different nature of the superpositions for each transfer eigenstate in the colorscale of Fig.~\ref{fig:4}.  

\section{Two-photon scattering properties~\label{sec:two}}

Here, we derive the two-photon scattering properties of the system using the same two approaches as in the previous section: First, in Section~\ref{subsec:effectivespintwophoton} we use the spin model derived in Eq.~\eqref{eq:spinmodel} to characterize the different two-polariton scattering eigenstates of the system in the $N\rightarrow \infty$ limit. This will provide insight on the different processes that can occur when two photons interact mediated by the non-linear emitter medium, such as elastic and inelastic collisions as well as scattering resonances \cite{BakkensenPhotonicQED}. Then, in Section~\ref{subsec:twophotonSmatrix}, we use the complete light-matter Hamiltonian of Eq.~\eqref{eq: complete4} to calculate the two-photon S-matrix that allows us to characterize the exact photonic output after the interaction of a non-zero bandwidth input with a finite number of emitters. 

An important difference with respect to the single-photon case is that the scattering of two photons only conserves the total energy but not the individual one. This results into a continuum of possible inelastic scattering outputs, where the frequency of each individual photon changes. However, in the $N\rightarrow \infty$ limit, an additional constraint appears, that is, that the total quasimomentum of the polaritons must be conserved. The later constraints certain inelastic outcomes and results in purely elastic outputs for some frequencies.

\subsection{Using the effective spin model~\label{subsec:effectivespintwophoton}}

\begin{figure*}[tb]
    \centering
    \includegraphics[width=0.95\textwidth]{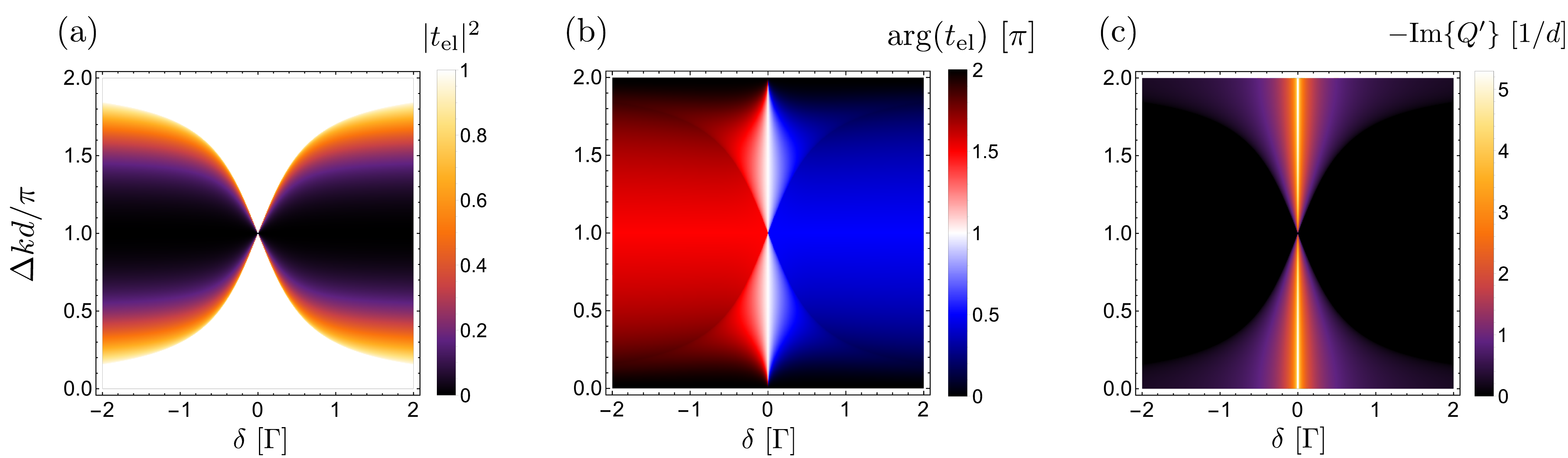}
    \caption{Two polariton scattering processes in the $N\rightarrow \infty$ and plane wave limit, when one polariton lies in the $+$ band and the other in the $-$ band, both having the same frequency detuning $\omega(q_+)=\omega(q_-)=\omega_0+\delta$. The parameters are $\Gamma_a=\Gamma_b=\Gamma/2$. (a) Elastic scattering output probability, as a function of $\delta$ and $\Delta k$. For small detunings $|\delta|\ll\Gamma$ (the specific range depends on $\Delta k$), a scattering resonance occurs (except for $\Delta kd=\pi$, where both polaritons have the same group velocity) and the output is purely elastic after the photons cross. (b) Non-linear phase acquired by the elastic output component, as a function of $\delta$ and $\Delta k$. We see that, for $|\delta|\ll \Gamma$, where the output is purely elastic, the phase depends linearly on the detuning and is $\pi$ on resonance: $\arg(t_\text{el})\equiv\varphi(\delta)=\pi+O(\delta)$ (except for $\Delta  kd=\pi$ and for $\Delta kd=\{0,2\pi \}$, where one polariton band disappears). (c) Imaginary part of the inelastic relative momentum, $-\text{Im}\{Q'\}$, as a function of $\delta$ and $\Delta k$. It is non-zero in the region where a scattering resonance occurs, leading to an elastic output after a finite binding time.}\label{fig:6}
\end{figure*}

In the limit $N\rightarrow \infty$, we can do an ansatz for the two-excitation eigenstates of the spin model of Eq.~\eqref{eq:spinmodel} using the two-excitation basis:
\begin{align}
    \label{eq:twopolbasis}\ket{K,Q}&\propto\sum_{n>m}e^{i(k_0+q_1)x_n}e^{i(k_0+q_2)x_m}\hat \sigma^\dagger_n\hat\sigma^\dagger_m\ket{g^{\otimes N}}=\nonumber\\
    &=\sum_{n>m}e^{iK x_c}e^{iQx_r}\hat \sigma^\dagger_n\hat\sigma^\dagger_m\ket{g^{\otimes N}}\ ,
\end{align}
where $K=2k_0+q_1+q_2$ represents the total quasimomentum of the polaritons, $Q=(q_1-q_2)/2$ the relative quasimomentum between them; $x_r=x_n-x_m$ is the relative distance between the excitations and $x_c=(x_n+x_m)/2$ their center of mass. Using this basis, we can write the two-polariton eigenstates as follows \cite{Schrinski_2022}:
\begin{equation}
\label{eq:twopolaritonansatz}
    \ket{\psi_{K,Q}}=\ket{K,Q}+t_\text{el}\ket{K,-Q} +t_\text{in}\ket{K,-Q'}\ \,
\end{equation}
where:
\begin{itemize}
    \item $t_\text{el}$ is the probability amplitude of elastic scattering, that is, the two polaritons cross ($q_1\leftrightarrow q_2$; $Q\rightarrow -Q$) without changing neither their individual single-particle frequencies nor their quasimomenta.
    \item $t_\text{in}$ is the probability amplitude of inelastic scattering, that is, the polaritons scatter into a new pair of polaritons with different quasimomenta and single-particle energies ($q_1,q_2\rightarrow q_2', q_1'$ ; $Q\rightarrow -Q'$).
\end{itemize}

The total momentum $K$ and energy $E_{K,Q}$ always remains conserved, i.e., $K=2k_0+q_1+q_2=2k_0+q'_2+q'_1$  and $E_{K,Q}=\omega(q_1)+\omega(q_2)=\omega(q_2')+\omega(q_1')$, where $\omega(q)$ is the Markovian single particle polariton dispersion as in Eq.~\eqref{eq:wq}. This means that, for every two polariton inputs $q_1,q_2$ associated with $K,Q$, we first need to find the corresponding degenerate relative quasimomentum $Q'$, associated with the inelastically scattered polariton pair $q'_1,q_2'$, which fulfills conservation of energy and total quasimomentum. The analytic solution for $Q'$ can be found in Appendix~\ref{subsec:spinmodel}. Then, we can find the elastic and inelastic scattering amplitudes, $t_\text{el}(K,Q)$ and $t_\text{in}(K,Q)$ by solving the time-independent Schrödinger equation: $\hat H_\text{eff}\ket{\psi_{K,Q}}=E_{K,Q}\ket{\psi_{K,Q}}$, with the ansatz of Eq.~\eqref{eq:twopolaritonansatz}, which is also shown in Appendix~\ref{subsec:spinmodel}. One then finds three different types of two-polariton processes, namely:
\begin{itemize}
    \item \emph{Elastic scattering}: for $|t_\text{el}|^2=1$ and $t_\text{in}=0$, where the input polaritons cross elastically, acquiring a phase given by $\mathrm{arg}(t_\text{el})$.
    \item \emph{Inelastic scattering}: for $|t_\text{el}|^2< 1$ and $t_\text{in}\neq 0$, with $\text{Im}\left\{Q'\right\}=0$. The input polaritons can scatter inelastically into a new pair of polaritons with different single particle energies.
    
    \item \emph{Scattering resonances}: for $|t_\text{el}|^2=1$ and $t_\text{in}\neq 0$, with $\text{Im}\left\{Q'\right\}< 0$. The non-zero imaginary part of $-Q'$ leads to a localized resonance: $\braket{x_c,x_r|K,-Q'}\propto e^{iKx_c}e^{-iQ'x_r}$. The input polaritons bind together for a finite time while crossing before resulting in a final elastic output with a non-linear phase-shift determined by $\varphi\equiv\arg(t_\text{el})$.
\end{itemize}

Let us now explore the situation of interest of this manuscript, that is, we consider two polaritons, one in the $+$ and the other in the $-$ branch, with an energy near resonant to the emitter's frequency. This is equivalent to consider a two-polariton input with energies $\omega(q_+)=\omega(q_-)=\omega_0+\delta$, with $\delta$ being the detuning with respect to the emitters' frequency. In Fig.~\ref{fig:6}, we plot the dependence of the elastic probability amplitude, $|t_\mathrm{el}|^2$
, its argument $\mathrm{arg}(t_\mathrm{el})$, and the imaginary quasimomentum $\text{Im}\left\{Q'\right\}$ as a function of $\delta$ and $\Delta k$ for an emitter array with $\Gamma_a=\Gamma_b$ \footnote{For $ \pi<\Delta kd<2\pi$ we choose $q_1\equiv q_-$ and $q_2\equiv q_+$; while for $ 0<\Delta kd<\pi$ we choose $q_1\equiv q_+$ and $q_2\equiv q_-$, as the slowest polariton has to be initially ahead.}. There, we observe different regions with distinct characteristics:
\begin{itemize}
    \item For any $\Delta k d\neq n\pi$, with $n\in\mathbb{Z}$, a scattering resonance appears around $\delta\approx 0$. The argument of the elastic scattering amplitude, $\varphi=\mathrm{arg}(t_\mathrm{el})$, that is, the phase acquired by photons due to their non-linear interaction mediated by the emitters, depends on the detuning, and scales linearly with $\delta$, for $|\delta|\ll\Gamma$. Importantly, this phase is additional to the single-photon phases that we calculate in Section~\ref{sec:single}, and it is strictly $\pi$ when the photons come in resonance, $\delta=0$, independent of the value of $\Delta k d$:   $\varphi(\delta)=\pi+O(\delta)$.
    
    \item For $\Delta k d=2 n\pi$, the $-$ polariton becomes uncoupled to the emitters (light-like) and its band disappears, as in Fig.~\ref{fig:2}(a). Therefore, in this limit, we cannot obtain a non-linear phase-shift.
    
    \item For $\Delta k d=(2 n+1)\pi$, both polaritons have the same group velocity, see Eq.~\eqref{eq:vqresonant}, and thus, for a finite bandwidth pulse, they will not cross. In this limit, there is no scattering resonance.

\end{itemize}

\subsection{Using the S-matrix~\label{subsec:twophotonSmatrix}}

The previous section considers the results for plane-wave inputs and infinitely many emitters. Here, we use the S-matrix formalism to calculate the two-photon response for finite emitter arrays and finite frequency band-widths. We start calculating the two-photon scattering matrix, $\hat S_2$, for a single unit cell. For that, we use the real space representation of the Hamiltonian $\hat H$, see Eq.~\eqref{eq: complete4}, and solve the time independent Schrödinger equation for a single emitter at position 0:
\begin{equation}
    \hat H \ket{\Psi_2}=(\omega_1+\omega_2)\ket{\Psi_2}\,,\label{eq:2pho}
\end{equation}
using the following wavefunction ansatz for the two-excitation subspace:
\begin{widetext}
\begin{equation}
    \ket{\Psi_2}=\left[\sum_{\alpha}\left( \int dx\  e_\alpha(x)\ \hat \sigma^\dagger \hat \alpha^\dagger_x + \int dx_1dx_2\  \phi_{\alpha\alpha}(x_1,x_2)  \frac{\hat \alpha^\dagger_{x_1} \hat \alpha^\dagger_{x_2} }{\sqrt{2}}  \right) + \int dxdy\ \psi_{ab}(x,y)\ \hat a^\dagger_x \hat b^\dagger_y \right]\ket{\text{vac}}\otimes\ket{g}\,.\label{eq:2photoan}
\end{equation}
\end{widetext}

Here,
\begin{itemize}
    \item $e_\alpha(x)$ represents the probability amplitude of having a single photon in the $\alpha$ channel while the emitter is excited.

    \item $\phi_{\alpha\alpha}(x_1,x_2)$ is the two photon probability amplitude when both photons propagate within the same mode $\alpha$. Due to its bosonic nature, it must satisfy: $\phi_{\alpha\alpha}(x_1,x_2)=\phi_{\alpha\alpha}(x_2,x_1)$.

    \item  $\psi_{ab}(x,y)$ is the probability amplitude that each of the photons are in the different channels.
    
\end{itemize}

To get the different components of the two-photon scattering matrix we solve equation Eqs.~\eqref{eq:2pho}-\eqref{eq:2photoan} for different input conditions at one side of the emitter, i.e., $\phi_{\alpha\alpha}(x_1<0,x_2<0)=\frac{1}{2\sqrt{2}\pi c_\alpha}\left[ e^{ik_\alpha(\omega_1)x_1}e^{ik_\alpha(\omega_2)x_2} + (x_1 \leftrightarrow x_2) \right]$ and $\psi_{ab}(x<0,y<0)=\frac{1}{2\pi \sqrt{c_a c_b}} e^{ik_a(\omega_1)x}e^{ik_b(\omega_2)y}$, and calculate the corresponding output at the other side, i.e., $\phi_{\alpha\alpha}(x_1>0,x_2>0)$ and $\psi_{ab}(x>0,y>0)$. Doing that, we arrive to the following two-photon S-matrix:
\begin{widetext}
\begin{align}\label{eq:2photonSmatrix}
    \nonumber
    \hat s_2=&\sum_{\alpha,\beta;\lambda,\nu}\int_{\omega_1>\omega_2} d\omega_1d\omega_2\ t_{\beta\alpha}(\omega_1) t_{\nu\lambda}(\omega_2)\ket{\omega_1\omega_2}_{\beta\nu}\bra{\omega_1\omega_2}_{\alpha\lambda} \\
    & -\frac{1}{\pi}\sum_{\alpha,\beta;\lambda,\nu}\int_{\omega_1>\omega_2} d\omega_1d\omega_2\ [t_{\beta\alpha}(\omega_1)-\delta_{\beta\alpha}][t_{\nu\lambda}(\omega_2)-\delta_{\nu\lambda}]\ket{B_E}_{\beta\nu}\bra{\omega_1\omega_2}_{\alpha\lambda}\ ,
\end{align}

\end{widetext}
for $\alpha,\beta,\lambda,\nu=\{a,b\}$, where $\ket{\omega_1\omega_2}_{\alpha\lambda}=\ket{\omega_2\omega_1}_{\lambda\alpha}$ is a two photon plane-wave state consisting of one photon of frequency $\omega_1$ in mode $\alpha$ and another of frequency $\omega_2$ in mode $\lambda$ (if $\alpha=\lambda$, the associated wavefunction has bosonic particle exchange symmetry):
\begin{align}
    \nonumber \ket{\omega_1\omega_2}_{\alpha\alpha}&=\frac{1}{2\sqrt{2}\pi c_\alpha} \int dx_1dx_2\big[ e^{ik_\alpha(\omega_1)x_1}e^{ik_\alpha(\omega_2)x_2}\\ 
    \nonumber&+e^{ik_\alpha(\omega_1)x_2}e^{ik_\alpha(\omega_2)x_1}\big]\frac{\hat \alpha_{x_1}^\dagger \hat \alpha_{x_2}^\dagger}{\sqrt 2}\ket{\text{vac}}\ ,\\ \ket{\omega_1\omega_2}_{ab}&=\frac{1}{2\pi \sqrt{c_ac_b}} \int dx dy\ e^{ik_a(\omega_1)x}e^{ik_b(\omega_2)y}\ \hat a_{x}^\dagger \hat b_{y}^\dagger\ket{\text{vac}}\ .
\end{align}

\begin{figure*}[tb]
    \centering
    \includegraphics[width=0.9\textwidth]{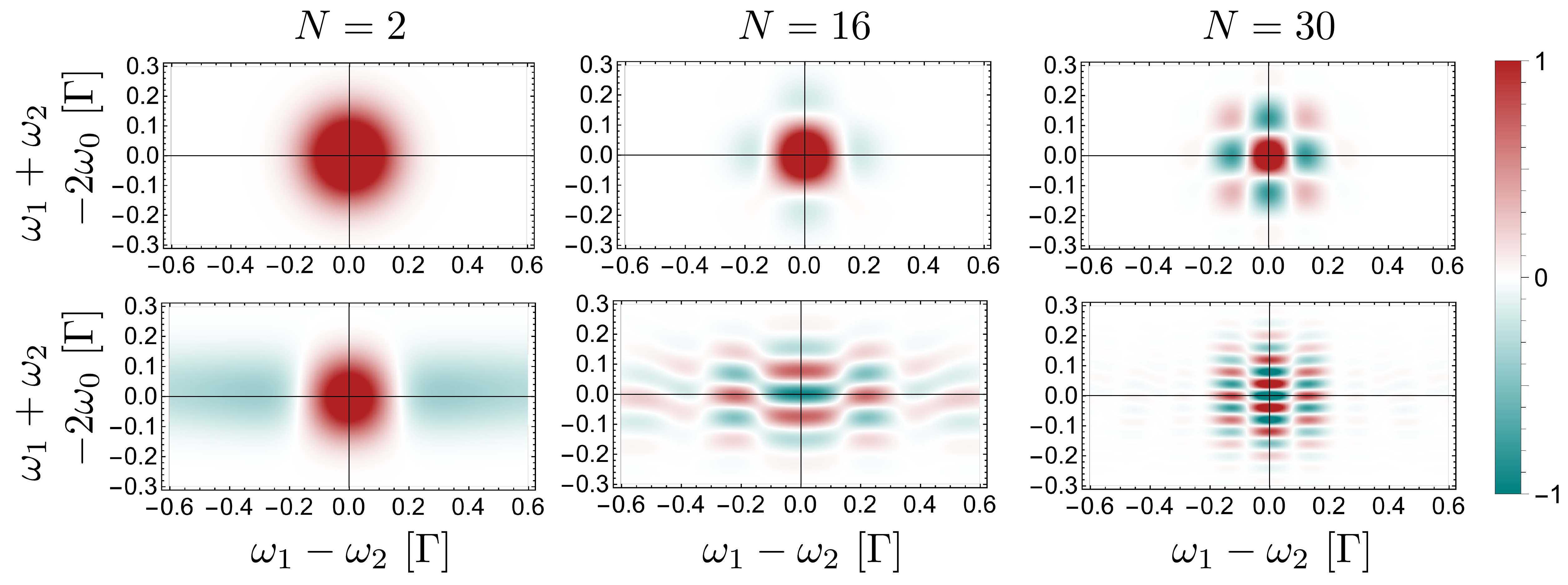}
    \caption{Input (top) and output (bottom) two-photon resonant wavepackets of bandwidth $ \sigma=0.05\Gamma$, for a setup with $\Delta kd=3\pi/2$ and different number of emitters $N$. The input consists of a photon in the $+$ superposition delayed $\tau$ with respect to the other photon in the $-$ superposition, such that they cross at the middle of the array. The colorscale indicates the real part of the wavefunction projected in the $\leftindex_{aa}{\bra{\omega_1\omega_2}}$ subspace, in arbitrary units. We see that, for $N=2$  there is an important inelastic component in the output. For $N=16$, the non-linear phase is achieved, but there is still an important inelastic contribution. For $N=30$ we achieve both a non-linear phase shift and an almost elastic ouput. }\label{fig:7}
\end{figure*}

The span of $\ket{\omega_1\omega_2}_{\alpha\lambda}$ for $\forall \alpha,\lambda$ and $\forall \omega_1,\omega_2\  |\ \omega_1>\omega_2$ forms an orthonormal complete basis for the two-photon sector, and thus satisfy $\leftindex_{\beta\nu}{\braket{\omega'_1\omega'_2|\omega_1\omega_2}}_{\alpha\lambda}=\delta_{\alpha\beta}\delta_{\lambda\nu}\delta(\omega_1-\omega_1')\delta(\omega_2-\omega_2')+\delta_{\alpha\nu}\delta_{\lambda\beta}\delta(\omega_1-\omega_2')\delta(\omega_2-\omega_1')$. In Eq.~\eqref{eq:2photonSmatrix}, the state $\ket{B_E}_{\beta\nu}$ represents a two-photon bound state of energy $E=\omega_1+\omega_2$, where one photon is in the $\beta$ mode and the other in the $\nu$ mode. In this state, the photons are bound together in the relative coordinate, and thus is decomposed as a continuum of plane-wave states, reading:
\begin{align}
    \ket{B_E}_{\beta\nu}&=\int_{\omega'_1>\omega'_2} d\omega'_1 d\omega'_2\ \delta(E-\omega_1'-\omega_2') \Big[\frac{1}{\frac{\Gamma}{2}-i(\omega_1'-\omega_0)}\nonumber\\
    &+ \frac{1}{\frac{\Gamma}{2}-i(\omega_2'-\omega_0)}\Big]\ket{\omega'_1\omega_2'}_{\beta\nu}\
\end{align}
where $\Gamma=\Gamma_a+\Gamma_b+\gamma$. 

With these definitions, we can understand that the first term in Eq.~\eqref{eq:2photonSmatrix} corresponds to the linear, elastic, uncorrelated transmission of each photon, and the second term to the non-linear, correlated transmission of the photons into a continuum of inelastic outputs where the single-photon energies are not conserved, only the total energy $E=\omega_1+\omega_2=\omega_1'+\omega'_2$.

The free propagation in the two-photon sector is given by:
\begin{align}
\nonumber
    &\hat P_{2}=\\
    &\sum_{\alpha,\lambda} \int_{\omega_1>\omega_2} d\omega_1d\omega_2\ e^{ik_\alpha(\omega_1)\frac{d}{2}}e^{ik_\lambda(\omega_2)\frac{d}{2}} \ket{\omega_1\omega_2}_{\alpha\lambda}\bra{\omega_1\omega_2}_{\alpha\lambda}\ .
\end{align}

With this formalism, we can calculate the two-photon response of an array of $N$ emitters against any two photon input. Let us conclude this section by showing what happens when we send a two-photon resonant input wavepacket of bandwidth $\sigma\ll\Gamma$, where one photon is prepared in the $+$ superposition and the other in the $-$ superposition, as in Eq.~\eqref{eqSM:eigenstates}. Inside the array, each polariton will acquire different delays for each emitter they encounter, $\tau_+$ and $\tau_-$, respectively, as in Eq.~\eqref{eq:delays}. Imprinting an initial delay
\begin{equation}
    \label{eq:initialdelay}
    \tau=(\tau_--\tau_+)\frac{N}{2}\ ,
\end{equation}
to the $+$ superposition, the photons will cross in the middle of the array, as schematically depicted in Fig.~\ref{fig:1}(c). In the top row of Fig.~\ref{fig:7}, we plot the projection into the $a$ channel of the real part of this input two-photon wavefunction for three different wavepackets prepared to cross in an array with $N=2$, $16$ and $30$ emitters, from left to right, respectively. We see how the wavepackets are Gaussians centered around $\omega_0$ with an oscillatory component coming from the phases induced by their temporal separation $\tau$.  In the bottom row, we plot the corresponding projection of their output two-photon wavepackets. There, we see how, for a small number of emitters, the only conserved quantity is the total energy of the photons and, as such, a continuum of inelastic outputs is available.  This results in the strong frequency correlations of the output wavepacket that we see for the $N=2$ case. However, as $N$ increases, such frequency correlations are filtered out due to the conservation of the total quasimomentum. As depicted, for $N=30$ emitters, the scattering is almost elastic and the resulting state just acquires a non-linear $\pi$-shift, plus some additional oscillatory phases due to the single-photon delays acquired in the transfer. This is consistent with the calculation in the $N\rightarrow \infty$ limit of Fig.~\ref{fig:6}. 

Let us finally emphasize again that the intuition behind this non-linear $\pi$-shift, summarized in Fig.~\ref{fig:1}(c), is that the chiral, multi-mode nature of the waveguide introduces a new degree of freedom, $\Delta k d$, which results in a different group velocity for the two resonant polaritons as they propagate within the emitter array. This allows to force situations in which they cross inside the non-linear medium, acquiring a $\pi$-shift less due to the saturable character of the emitters.

\section{Gate mechanism and performance~\label{sec:gate}}

\subsection{Ideal mechanism}
 
 Now, we show how to use the ingredients discussed in the previous sections to design  a conditional gate between photonic qubits, in particular, a CZ one defined by:
\begin{align}\label{eq:gateoperation}
    \nonumber
    \ket{0}_a\otimes \ket{0}_b
    &\rightarrow \,\, \ket{0}_a\otimes \ket{0}_b\,, \\
    \nonumber
   \ket{0}_a\otimes \ket{1}_b
    &\rightarrow \,\, \ket{0}_a\otimes \ket{1}_b\,, \\
    \ket{1}_a\otimes \ket{0}_b
    &\nonumber \rightarrow \,\, \ket{1}_a\otimes \ket{0}_b\,, \\
    \ket{1}_a\otimes \ket{1}_b
    &\rightarrow \,\, e^{i\pi}\ket{1}_a\otimes \ket{1}_b\,.
\end{align}

For that, one must first define: i) a suitable qubit encoding; ii) two distinguishable subsystems where these qubits can propagate. In our case, we can define the gate using two different qubit encodings, as schematically represented in Fig.~\ref{fig:Encodings}(b-c):
\begin{itemize}
    \item A photon-number encoding in which the presence or absence of a photon in channel $\alpha=\{a,b\}$ represents the logical 1 and 0, respectively, i.e., $\ket{1}_\alpha \equiv \ket{\omega_0}_\alpha$ and $\ket{0}_\alpha \equiv \ket{\mathrm{vac}}_\alpha$.
    
    \item A dual-rail encoding in which a non-interacting  auxiliary channel $\alpha'$ is added for each main interacting channel $\alpha$, such that $\ket{1}_\alpha$ and $\ket{0}_\alpha$ represents the presence of a photon in the main or auxiliary channel, respectively: $\ket{1}_\alpha\equiv\ket{\omega_0}_\alpha\otimes\ket{\text{vac}}_{\alpha'}$ and $\ket{0}_\alpha\equiv\ket{\text{vac}}_\alpha\otimes\ket{\omega_0}_{\alpha'}$.
\end{itemize}

Irrespective of the choice of the encodings, the gate architecture of Fig.~\ref{fig:Encodings}(b-c) is composed by three elements:
\begin{itemize}
    \item An initial and final beam splitter operation $\hat t$ that transforms the resonant photons propagating in the main $a$ ($b$) channel into the $+$ ($-$) transfer eigenstate and vice versa, see Fig.~\ref{fig:Encodings}(a):
    \begin{align}   
\label{eq:BeamSplitter}
    \nonumber \ket{\omega_0}_a&\xrightarrow{\hat{t}}\ket{\omega_0}_+\xrightarrow{\hat{t}}\ket{\omega_0}_a\\ \ket{\omega_0}_b&\xrightarrow{\hat{t}}\ket{\omega_0}_-\xrightarrow{\hat{t}}\ket{\omega_0}_b\,.
\end{align}

    These transformations are required since the photons propagating in the original modes are not the transfer eigenstates of the system acquiring the desired phases. Thus, we need to convert them before entering the array, and then transform them back into the original qubit modes. The implementation of such beam splitter transformation will also depend on the particular realization. Nonetheless, in Appendix~\ref{subsec:beam} we provide a possible implementation using a single linear scatterer coupled to the waveguides, which will be the one we use for the calculations. Finally, let us note that, since the transfer eigenstates have an intricate frequency dependence, see Eq.~\eqref{eqSM:eigenstates}, this transformation will only be exact when the pulse width of the photonic qubit goes to zero, i.e., $\sigma\rightarrow 0$.  
    
    \item Apart from the two beam splitter transformations, the key ingredient is the two-level emitter array to induce the conditional phase shift between the two polaritons, like we demonstrate in Fig.~\ref{fig:7}.
   
\end{itemize}

Then, the final resulting output of that architecture can be described by a global S-matrix $\hat{G}$, including the two beam splitter operations $\hat t$ and the $N$-emitters array:
\begin{equation}
    \hat{G}=\hat{t}\ \hat{S}^N\ \hat{t}\ ,
\end{equation}
where $\hat{S}$ is the unit cell S-matrix from Eq.~\eqref{eq:Sunitcell} with parameters  $\Gamma_a=\Gamma_b$. Since we assume the beam splitters to be linear elements, the global S-matrix within the $n$-photon sector can always be written as $\hat{G}_n=\hat{t}\ \hat{S}_n^N\ \hat{t}$. The results of sections~\ref{sec:single}-\ref{sec:gate} demonstrate that: i) in the limit of small frequency band-width $\sigma\rightarrow 0$; ii) resonant input; iii) no other losses ($\gamma=0$), and iv) large and even number of emitters $N\rightarrow\infty$, the architecture depicted in Fig.~\ref{fig:Encodings}(b-c) constitutes a CZ gate operation as in Eq.~\eqref{eq:gateoperation}.

\subsection{Fidelity calculation for relevant experimental parameters}

In general, the aforementioned ideal conditions (i-iv) are not met, and thus the behavior of the gate will be different than the ideal one described in Eq.~\eqref{eq:gateoperation}.  To characterize the gate performance in realistic conditions, we use the following definitions for the single and two-photon input pulses:
\begin{itemize}
    \item $\ket{\Psi_{\text{in},\sigma}}_{a(b)}$ are single-photon resonant Gaussian pulses of bandwidth $\sigma$ entering from the main channel $a$ ($b$):
    \begin{align}
\label{pulsedefinsph}
\ket{\Psi_\mathrm{in,\sigma}}_{a(b)}&=\frac{1}{\sqrt{\sigma\sqrt{2\pi}}}\int d\omega\ e^{-\frac{(\omega-\omega_0)^2}{4\sigma^2}}\ket{\omega}_{a(b)}\, .
\end{align}
    \item $\ket{\Psi_{\text{in},\sigma}}_{ab}$ is a two-photon resonant Gaussian pulse where one photon comes from the $a$ channel and the other from the $b$ channel, with a form defined by a product state of $\ket{\Psi_{\text{in},\sigma}}_{a}$ and $\ket{\Psi_{\text{in},\sigma}}_{b}$, as in Eq.~\eqref{pulsedefinsph}. A delay $\tau$, as in Eq.~\eqref{eq:initialdelay}, is to be imprinted to the $a$-channel photon (which will map into the $+$ polariton via the beam splitter), such that the pulses cross at the middle of the array, as schematically shown in Fig.~\ref{fig:1}(c).
\end{itemize}

The exact outputs given these inputs are then given by $\ket{\Psi_{\text{out,}\sigma}}=\hat G \ket{\Psi_{\text{in,}\sigma}}$, which we will benchmark against the following ideal single and two-photon outputs:
\begin{widetext}
\begin{align}
\label{eqSM:pulsedefidealsph}
\ket{\Psi_\mathrm{out\  (id),\sigma}}_{b[a]}&=[(-1)^N]\frac{e^{i k_0 Nd}}{\sqrt{\sigma\sqrt{2\pi}}}\int d\omega\ e^{-\frac{(\omega-\omega_0)^2}{4\sigma^2}}e^{i(\omega-\omega_0)N\tau_{-[+]}}\ket{\omega}_{b[a]}\,,\\
\ket{\Psi_\mathrm{out\  (id),\sigma}}_{ab}&=e^{i\varphi}(-1)^N\frac{e^{2i k_0 Nd}}{\sigma\sqrt{2\pi}}\int d\omega_1 d\omega_2\ e^{-\frac{(\omega_1-\omega_0)^2}{4\sigma^2}}e^{-\frac{(\omega_2-\omega_0)^2}{4\sigma^2}}e^{i(\omega_1-\omega_0)(N\tau_{+}+\tau)}e^{i(\omega_2-\omega_0)N\tau_{-}}\ket{\omega_1\omega_2}_{ab}\ .
\label{eqSM:pulsedefideal}
\end{align}
\end{widetext}
which are defined assuming: 
\begin{itemize}
    \item A perfect transformation of the original $a$ ($b$) channel photons into the transfer eigenstates of the array $+$ ($-$) and back, via the beam splitter operation.
    \item A perfect transmission of the photons through the array, where the original $a$-channel photon (transformed into a $+$ superposition), acquires a $\pi$ phase and a $\tau_+$ delay for each unit cell of the array; and the original $b$-channel photon (transformed into a $-$ superposition), acquires no phase and a $\tau_-$ delay for each unit cell of the array, as given by Eq.~\eqref{eq:delays}.
    \item The two-photon ideal output consists of the product state of single photon outputs with only a phase shift $\varphi=\pi$ of the two-photon input trough non-linear processes, thus without any spectral entanglement.
\end{itemize}

\begin{figure}[tb]
    \centering
    \includegraphics[width=0.83\columnwidth]{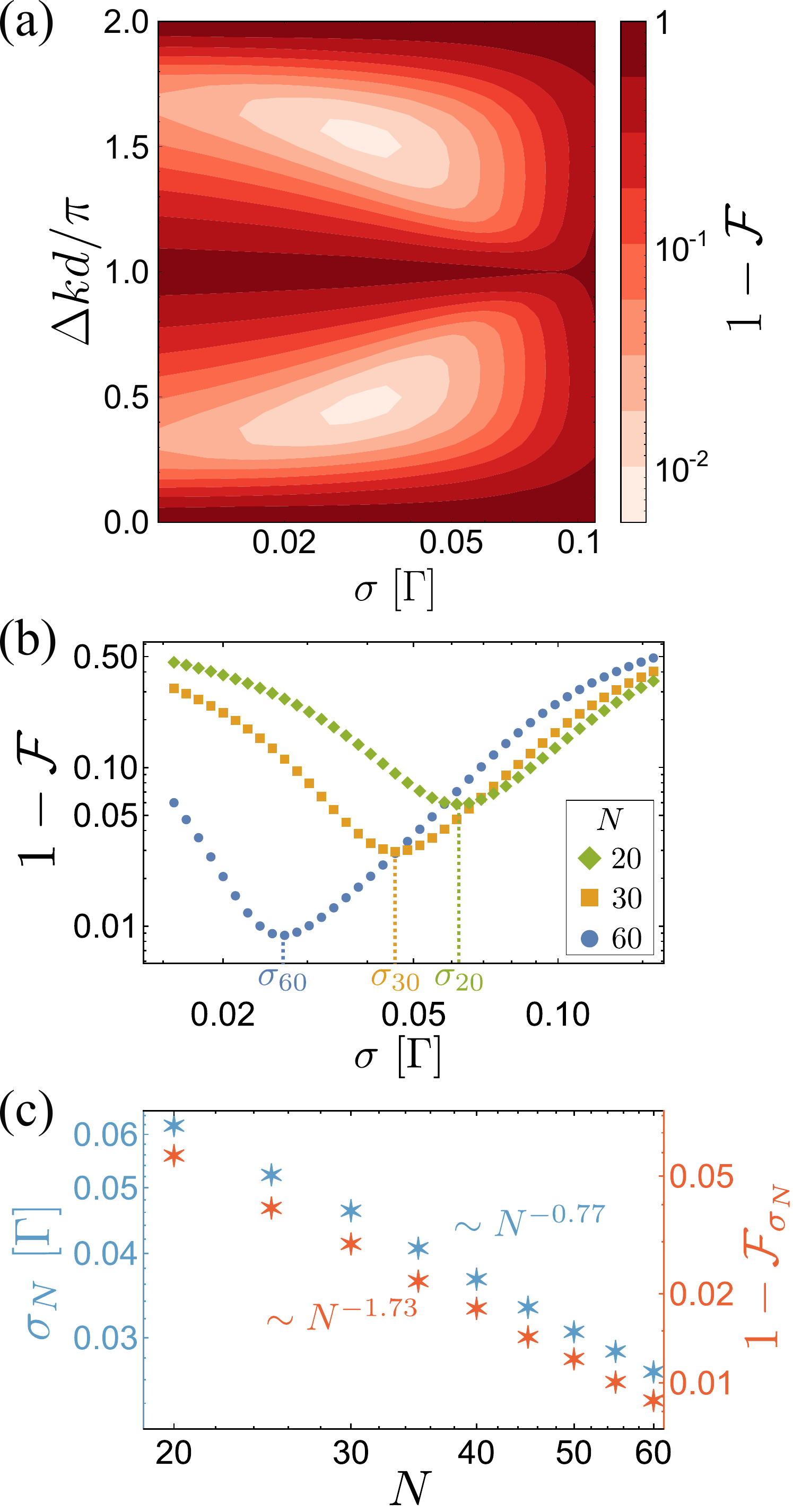}
    \caption{(a) Gate infidelity dependence as a function of the pulse width $\sigma$ and multi-mode phase difference $\Delta k d$ for $N=60$ emitters. (b) Infidelity of the gate as a function of the input bandwidth, for different number of emitters in the array, $N$, as in legend. For every $N$, there is an optimal bandwidth $\sigma_N$, such that the pulse is narrow in frequency but also narrow enough in time to fit in the array. (c) Optimal bandwidth $\sigma_N$ and infidelity as a function of the number of emitters in the array, $N$. As $N\rightarrow \infty$, the optimal bandwidth decreases monotonically as a power law: a larger array fits a smaller bandwidth pulse in its bulk. Parameters: In all figures, we fix the initial photon delay so that the photons cross at the middle of the emitter array. In panels (b) and (c) we fix $\Delta k d=3\pi/2$.} 
    \label{fig:9}
\end{figure}

Using the previous definitions, we calculate the fidelity of the gate as follows~\cite{Schrinski2022PassiveEmitters}:
\begin{align}\label{eq:fidelity}
    \mathcal{F}&=\frac{1}{16}\Big| 1\ +\  \leftindex_a{\braket{\Psi_{\text{out\ (id)},\sigma}|\Psi_{\text{out},\sigma}}}_a\  +\ \leftindex_b{\braket{\Psi_{\text{out\ (id)},\sigma}|\Psi_{\text{out},\sigma}}}_b\ \nonumber\\
    &+\leftindex_{ab}{\braket{\Psi_{\text{out\ (id)},\sigma}|\Psi_{\text{out},\sigma}}}_{ab}   \Big|^2\ .
\end{align}
This formula will account for any deviations of the ideal transfer such as imperfect preparation of the transfer scattering states and deformations of the wavepackets 
as finite pulse widths probe the non-linear region of the polariton dispersion; deviations from the scattering resonance region in Fig.~\ref{fig:6}; or the spectral correlations induced by small number of emitters $N$. Equipped with this formalism, let us now analyze the role of the different relevant experimental parameters in the gate performance.

We start by plotting in Fig.~\ref{fig:9}(a) the dependence of the infidelity of the gate, $1-\mathcal{F}$, as a function of both the pulse input band-width $\sigma$ and the effective multi-mode phase difference $\Delta k d$ for a fixed number of emitters $N=60$~\footnote{We see that there is a symmetry under the transformation $\Delta kd \rightarrow 2\pi- \Delta kd$. This is because, doing this, the dispersion relation just interchanges $v(q_+) \leftrightarrow v(q_-)$.}. From there, we can make a first important observation, that is, not all multi-mode phase differences $\Delta k d$ are valid to obtain the gate. For example, both $\Delta kd\sim \pi$ and $\Delta kd\sim 0\ (2\pi)$ situations lead to very large errors irrespective of the pulse width. The reason is that in the former case ($\Delta kd\sim \pi$) the two resonant polaritons have the same velocity, and thus they can not cross to obtain the conditional phase shift. In the other scenario, $\Delta kd\sim 0\ (2\pi)$, the gate fails because one polariton band effectively disappears and, thus, one can not make the propagating photons overlap as they cross the chain. The minimum error occurs for $\Delta k d \sim 3\pi/2,\pi/2$. All these results are in accordance to the predictions in the $N\rightarrow\infty$ and $\sigma\rightarrow0$ limit of Fig.~\ref{fig:6}.
 
 Beyond this dependence on $\Delta k d$, Fig.~\ref{fig:9}(a) also points to an optimal band-width which minimizes the error of the gate. The intuition behind it is that, on the one hand, one needs narrow wavepackets ($\sigma\ll \Gamma$) in frequency/momentum space to be close to the plane-wave limit where performance is maximized. On the other hand, the spatial shape of the wavepacket, proportional to $1/\sigma$, has to fit within the finite size of the array, $N d$. This trade-off is more clear in Fig.~\ref{fig:9}(b) where we precisely plot the infidelity for $\Delta k d=3\pi/2$, for several array sizes in different colors. There, we observe how indeed for each system size, there is an optimal band-width, that we denote as $\sigma_N$, for which the error is minimum. In Fig.~\ref{fig:9}(c), we plot both the scaling of the optimal band-width and infidelity with the number of emitters, showing how it approaches to the ideal performance as $\sigma_N\sim1/N^{0.77}$, $1-\mathcal{F}_{\sigma_N}\sim1/N^{1.73}$ .
 
\begin{figure}[tb]
    \centering
    \includegraphics[width=0.75\columnwidth]{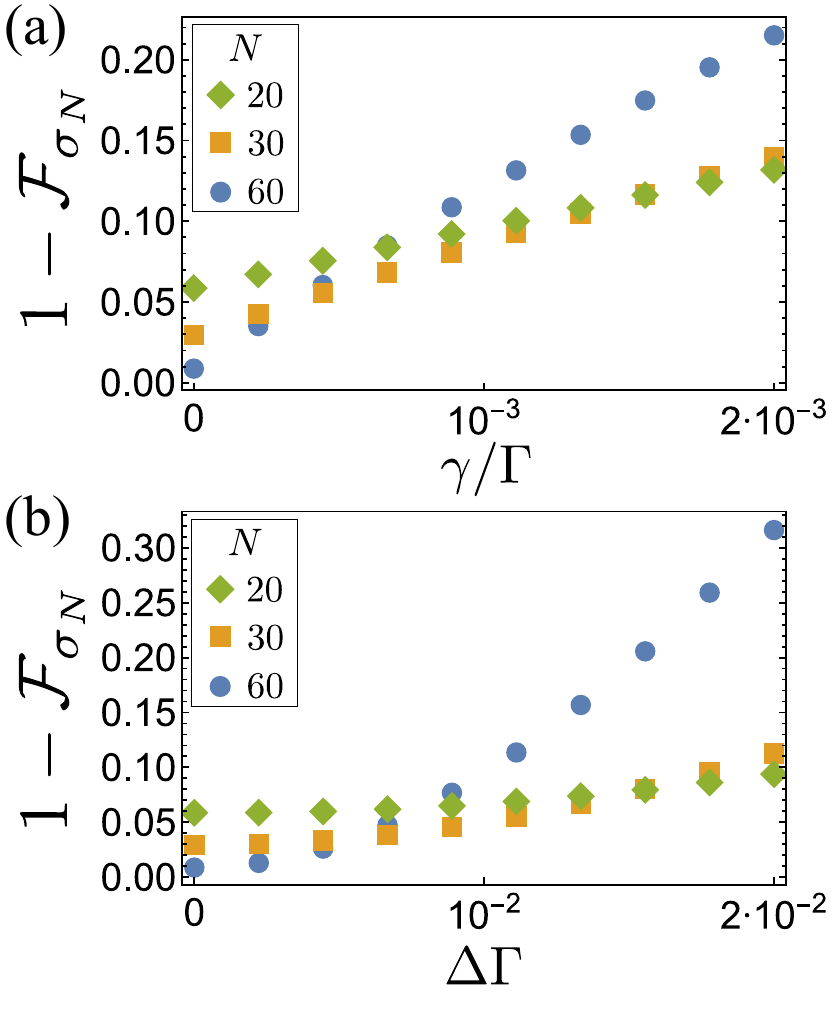}
    \caption{Infidelity dependence as a function of the decay into other channels $\gamma/\Gamma$ (a) and coupling imperfections $\Delta\Gamma$ (b) for different number of emitters, as depicted in the legend. The calculations are made for $\Delta k d=3\pi/2$, fixing the rest of the parameters such as time-delay or pulse-width to obtain the optimal performance of the gate.} 
    \label{fig:10}
\end{figure}

Apart from this fundamental trade-off between the pulse band-width and system size inherent to the protocol, there are some other experimental imperfections that always affect the performance of photonic quantum gates in these setups, and which will also impact ours in a similar fashion. In particular, the decay into other non-guided photonic modes, i.e., $\gamma\neq 0$, leads to a decrease of the fidelity which is more significant the more emitters there are, as shown in Fig.~\ref{fig:10}(a). Note, however, that in the dual rail encoding situation part of these errors can be corrected through post-processing, at the expense of making the gate probabilistic. Another potential source of error is the imperfect coupling of the emitters to the two modes, that one can characterize by an adimensional parameter, $\Delta \Gamma=(\Gamma_a-\Gamma_b)/\Gamma$, and which leads to a decrease of fidelity that we show in Fig.~\ref{fig:10}(b).

\section{Conclusions~\label{sec:conclusions}}

Summing up, we propose how to engineer a passive quantum phase gate between co-propagating photons using only two-level emitter non-linearities. The key ingredient is the chiral coupling of an emitter array to two co-propagating guided modes with different energy dispersion, as it appears naturally in the edges of two-dimensional topological lattices. By studying the single and two-photon scattering response, we show how the non-linear interactions in these systems can be harnessed to obtain a conditional $\pi$-shift without spectral entanglement. We optimize the performance of the gate as a function of relevant experimental parameters, such as pulse band-width and number of emitters, showing it can reach high fidelities for large number of emitters. 

Apart from the possibility to engineer quantum gates in topological photonic scenarios, our designs can be implemented in other setups where these chiral multi-mode setups can be engineered such as spin-orbit coupled atoms or quantum dots coupled to optical fibers and photonic crystal waveguides, respectively. An interesting outlook of this work is precisely to propose realistic implementations of these setups beyond the schematic designs of Fig.~\ref{fig:Encodings}(d-f). These setups could be integrated into more complex photonic quantum circuits as strong few-photon non-linearities enabling universal manipulation of quantum information~\cite{obrien_photonic_2009,Lodahl_2018,gonzalez-tudela_lightmatter_2024}.

\begin{acknowledgments}
 The authors acknowledge support from the Proyecto Sin\'ergico CAM 2020 Y2020/TCS-6545 (NanoQuCo-CM), the CSIC Research Platform on Quantum Technologies PTI-001 and from Spanish projects PID2021-127968NB-I00 funded by MICIU/AEI/10.13039/501100011033/ and by FEDER Una manera de hacer Europa, and TED2021-130552BC22 funded by MICIU/AEI /10.13039/501100011033 and by the European Union NextGenerationEU/ PRTR,
respectively. AGT also acknowledges a 2022 Leonardo Grant for Researchers and Cultural Creators,
and BBVA Foundation. TR further acknowledges support from the Ram\'{o}n y Cajal program RYC2021-032473-I, financed by MCIN/AEI/10.13039/501100011033 and the European Union NextGenerationEU/PRTR. TLY acknowledges funding from the Ministry of Science, Innovation and Universities of Spain FPU22/02005. The codes to reproduce the results of the manuscript are openly available \cite{tomas_levy_yeyati_2024}.
\end{acknowledgments}

\bibliographystyle{unsrt}
\bibliography{references,referencesAlex}

\appendix
\section{Two polariton scattering in the infinite emitters limit}\label{subsec:spinmodel}
We follow the analysis done in Ref.~\cite{Schrinski_2022} for 
a single bidirectional waveguide. First, we find the pair of polaritons $q_1'$,$q_2'$, that have the same energy and total quasimomentum as a given pair of  polaritons, $q_1$,$q_2$. For this we rewrite them as total and relative quasimomentum,  $K=2k_0+q_1+q_2=2k_0+q_1'+q_2'$ and $Q=(q_1-q_2)/2$, $Q'=(q'_1-q'_2)/2$. This way, we only need to solve the equation
\begin{equation}
\omega(q_1)+\omega(q_2)=\omega(q_1')+\omega(q_2')\ ,
\end{equation}
for $Q'$, where $\omega(q)$ is the Markovian polariton dispersion of~\eqref{eq:wq}. Changing variables, we arrive to the following expression:
\begin{widetext}
\begin{align}
\nonumber
&\cos\left(Q'd\right) =\\
&\frac{2r\cos^2\left[\left(\frac{\Delta k}{2}+\kappa\right)d\right]\sin\left[\left(\frac{\Delta k}{2}-\kappa\right)d\right]+\cos\left(Qd\right)\left[(r-1)\sin(2\kappa d)-(r+1)\sin(\Delta k d)\right] + 2\cos^2\left[\left(\frac{\Delta k}{2}-\kappa\right)d\right]\sin\left[\left(\frac{\Delta k}{2}+\kappa\right)d\right]}{2(r-1)\sin(\kappa d)\left[\cos(Q d)\cos\left(\frac{\Delta k}{2} d\right)-\cos(\kappa d)\right] + 2(r+1)\sin\left(\frac{\Delta k}{2}d\right)\left[\cos\left(\frac{\Delta k}{2} d\right)-\cos(Q d)\cos(\kappa d)\right]}\ ,
\end{align}
\end{widetext}
where $r=-\Gamma_b/\Gamma_a$ and $\kappa = K/2-k_0$. Once we have the degenerate quasimomentum, we solve the time-independent Schrödinger equation: $\hat H_\text{eff}\ket{\psi_{K,Q}}=E_{K,Q}\ket{\psi_{K,Q}}$, with the ansatz of Eq.~\eqref{eq:twopolaritonansatz}. To do it, we just need to know the action of $\hat H_\text{eff}$ in the basis $\ket{K,Q}$ of Eq.~\eqref{eq:twopolbasis},  which is:
\begin{align}
\nonumber
\hat H_\text{eff}\ket{K,Q}=&\ E_{K,Q}\ket{K,Q}\\
    &+\Gamma_a\left[ i-f(Q,k_a-K/2)\right]\ket{K,k_a-K/2} \nonumber \\
    &+\Gamma_b\left[ i-f(Q,k_b-K/2)\right]\ket{K,k_b-K/2} \ ,
\end{align}
where $f(x,y)=\sin(xd)/[\cos(xd)-\cos(yd)]$. Doing this, we arrive to a linear system of equations for the elastic and inelastic probability amplitudes, depending on the input polaritons $q_1,q_2$, associated to the total, $K$, and relative, $Q$, quasimomenta:

\begin{widetext}
    \begin{align}
    1+i f(Q,k_a-K/2)+t_\text{el}[1-if(Q,k_a-K/2)]+t_\text{in}[1-if(Q',k_a-K/2)]=0\ ,\\
    1+i f(Q,k_b-K/2)+t_\text{el}[1-if(Q,k_b-K/2)]+t_\text{in}[1-if(Q',k_b-K/2)]=0\ .
\end{align}
\end{widetext}

\section{Beam splitter operating regime for an individual scatterer~\label{subsec:beam}}


Let us make an interesting observation regarding the form of the single-photon S-matrix for a single emitter, see Eq.~\eqref{eq:s1Matrix}, that is, that it defines a general beam splitting operation between the original waveguide modes. This means that controlling the ratio $\Gamma_a/\Gamma_b$, one can obtain different superpositions between the $a$ and $b$ channels.  In particular, it can be used to transform single photons on resonance in the $a$ $(b)$ modes into the $+$ $(-)$ transfer eigenstates and, likewise, for the inverse operation, transforming $+$ $(-)$ transfer eigenstates into purely $a$ $(b)$ modes. For that, one must find the conditions such that $t_{aa}(\omega_0)=t_{ba}(\omega_0)=-t_{bb}(\omega_0)$, which occurs when $\Gamma_a=(2+\sqrt{2})\Gamma/4$ and $\Gamma_b=(2-\sqrt{2})\Gamma/4$,  being $\Gamma=\Gamma_a+\Gamma_b$. Denoting as $\hat{t}$ to the $\hat{s}_1$ matrix obtained using these parameters, it acts as follows for resonant photons:
\begin{align}  
    \label{eqSM:beamsplitter}
    \nonumber \ket{\omega_0}_a&\xrightarrow{\hat{t}}-\ket{\omega_0}_+\xrightarrow{\hat{t}}\ket{\omega_0}_a\\ \ket{\omega_0}_b&\xrightarrow{\hat{t}}-\ket{\omega_0}_-\xrightarrow{\hat{t}}\ket{\omega_0}_b\ .
\end{align}
This provides a constructive way of preparing the transfer eigenstates of the array $\ket{\omega_0}_\pm$ from single photons propagating in the original waveguide channels, by placing such a beam splitter at the beginning of the array; and then retrieving back single channel modes after they propagate through the array, placing one at the end. Let us make some remarks about this operation:
\begin{itemize}
   
    \item To maintain the linearity of the operation in the multi-photon regime, the scatterer performing the beam-splitting must have a linear spectrum, e.g., cavity.

    \item Apart from creating superpositions between the original $a$ and $b$ channels, the $\hat{t}$-transformation also induces a different delay phase within the photon propagation in the $a/b$ channels. Up to first order in $(\omega-\omega_0)/\Gamma$, the delays imprinted to the $a/b$ photons due to the two beam splitters are given by:
    \begin{align}
        \nonumber
        \tau_a&=(4+2\sqrt{2})/\Gamma\,,\\
        \tau_b&=(4-2\sqrt{2})/\Gamma\,.\label{eq:delayab}    \end{align}
    
    \item Finally, the transformation into a transfer eigenstate is only exact for photons coming exactly on resonance $\omega=\omega_0$. For finite pulses with frequency band-widths $\gtrsim \Gamma$, the transformation will have an error that we take into account in Section~\ref{sec:gate}.

\end{itemize}

\end{document}